\documentclass[aps,showpacs,preprint,amsmath,floatfix]{revtex4}

\usepackage{graphicx}
\usepackage{epsfig}
\usepackage{dcolumn}
\usepackage{bm}

\begin{document}

\title{Vus and neutron beta decay}

\author{A.~Garc\'{\i}a}

\affiliation{
Departamento de F\'{\i}sica.\\
Centro de Investigaci\'on y de Estudios Avanzados del IPN.\\
A.P. 14-740.\\
M\'exico, D.F., 07000. M\'EXICO.
}

\author{G.~S\'anchez-Col\'on}

\email[]{gsanchez@mda.cinvestav.mx}

\affiliation{
Departamento de F\'{\i}sica Aplicada.\\
Centro de Investigaci\'on y de Estudios Avanzados del IPN.\\
Unidad M\'erida.\\ A.P. 73, Cordemex. \\
M\'erida, Yucat\'an, 97310. M\'EXICO.
}

\date{\today}

\begin{abstract}
We discuss the effect of the recent change of $V_{\rm us}$ by
three standard deviations on the standard model predictions for
neutron beta decay observables. We also discuss the effect the
experimental error bars of $V_{\rm us}$ have on such
predictions. Refined precision tests of the standard model will
be made by a combined effort to improve measurements in neutron
beta decay and in strangeness-changing decays. By itself the
former will yield very precise measurements of $V_{\rm ud}$ and
make also very precise predictions for $V_{\rm us}$.
\end{abstract}


\maketitle

\section{\label{secone}Introduction}

The precision measurements of the decay rate R and the
electron-asymmetry $\alpha_e$ in neutron beta decay
(n$\beta$d)~\cite{pdg06} and their further improvements in a near
future provide an excellent opportunity to test the standard
model (SM)~\cite{review} and even to establish deviations signaling
new physics. However, the predictions for these observables are
afflicted by our current inability to compute reliably the
Cabibbo-Kobayashi-Maskawa (CKM) matrix element $V_{\rm ud}$ and
the leading form factor ratio $\lambda=g_1/f_1$. Both are better
handled as free parameters to be determined from experiment. The
theoretical predictions are then confined to a region in the
$(\alpha_e, R)$ plane or equivalently in the $(\lambda, R)$
plane where the SM is expected to remain valid within a certain
confidence level (CL), say $90\%$. This region may be referred to
as the standard model region (SMR). At first, it may look as if
the predictions of SM are severely limited by the experimental
situation of $R$ and $\alpha_e$. However, this is not the case.

In a previous paper~\cite{garcia02} we showed that the SMR is
determined by the validity of the formulas predicted by the SM
for the observables in n$\beta$d and of the CKM unitarity. The
size of the SMR depends on the theoretical uncertainties of such
formulas and the experimental values of $V_{\rm us}$ and $V_{\rm
ub}$. Since such uncertainties in $R$ and $\alpha_e$ are
substantially smaller than their experimental error bars, a much
more narrow SMR can be predicted even when $V_{\rm ud}$ and
$\lambda$ remain as free parameters. The predictions of SM are
then greatly improved and it is these ones that are meaningful to
compare with the measured $R$ and $\alpha_e$.

Nevertheless, such predictions are indeed affected by the
experimental values of $V_{\rm ub}$ and $V_{\rm us}$. The former
is quite precise already and its changes do not produce
perceptible changes in the SMR. However, changes in $V_{\rm us}$
do produce important changes in the position and size of the SMR.
It is the purpose of this paper to extend the analysis of
Ref.~\cite{garcia02} and discuss the dependence of the SMR on the
value of $V_{\rm us}$. This has become more pressing since
recently~\cite{pdg06} its experimental value increased by three
standard deviations from the value available for the analysis
of~\cite{garcia02}.

In Sec.~\ref{sectwo} we shall review the SM formulas for
n$\beta$d observables and the method to determine the predicted
SMR. In Sec.~\ref{secthree} we shall determine the changes in
the SMR corresponding to the new value of $V_{\rm us}$. We shall
also determine its position allowing for variations of up to
three standard deviations of the present $V_{\rm us}$. The role
of $V_{\rm us}$ has another aspect, its precision affects
importantly the size of the SMR. This will be studied in
Sec.~\ref{secfour}. A complementary analysis comes from the fact
that precise measurements of $R$ and $\alpha_e$ will produce a
precise determination of $V_{\rm ud}$. Assuming the validity of
the unitarity of the CKM matrix, then n$\beta$d can make quite
precise predictions for $V_{\rm us}$. We shall go into them in
Sec.~\ref{secfive}. The last section is reserved for discussions
and conclusions.

\section{\label{sectwo}Determination of the standard model region}

The SM predicts for the decay rate of n$\beta$d the expression

\begin{equation}
R(10^{-3}\,{\rm s}^{-1}) =
|V_{\rm ud}|^2(0.1897)(1 + 3\lambda^2)(1 + 0.0739 \pm 0.0008)
\label{R}
\end{equation}

\noindent
at the level of a precision of $10^{-4}$. $V_{\rm ud}$ and
$\lambda$ appear as free parameters. The detailed derivation of
Eq.~(\ref{R}) is found in Ref.~\cite{garcia01}. The main source
of uncertainty in~(\ref{R}) is the model dependence of the
contributions of $Z^0$ to the radiative corrections. A very
conservative estimate is $\pm 0.0008$~\cite{marciano86}. If one
assumes dominance of the $A_1$ resonance~\cite{wilkinson94} this
uncertainty becomes the uncertainty of such an approximation and
then in Eq.~(\ref{R}) it can be estimated to be somewhat less
than $\pm 0.0002$. Other uncertainties as in the values of the
induced weak magnetism and pseudo-tensor form factors can be
shown to contribute to $10^{-5}$ or less. Eq.~(\ref{R}) has also
been discussed in Ref.~\cite{czarnecki04}, where it was referred
to as the master formula. Although presented in a somewhat
different form, one can readily verify that the result of this
reference confirms Eq.~(\ref{R}).

At the $10^{-4}$ level the SM predicts for the electron-asymmetry
the expression~\cite{garcialuna06}

\begin{equation}
\alpha_e =
\frac{-0.2089\times 10^{-3} + 0.2763\lambda - 0.2772\lambda^2}
{0.1897 + 0.5692\lambda^2}.
\label{alphae}
\end{equation}

\noindent
We have chosen a negative sign for $\lambda$ to conform with the
convention of~\cite{pdg06}. The important remark here is that
there is no theoretical uncertainty in $\alpha_e$ at this level
of precision. The reason for this is that the uncertainty
introduced by $Z^0$ is common to the numerator and denominator
of $\alpha_e$ and cancels away at the $10^{-4}$ level. It must
be stressed that $\alpha_e$ depends only on $\lambda$, so that
the experimental determination of $\lambda$ is independent of
$V_{\rm ud}$.

The analysis that leads to Eq.~(\ref{alphae}) can be extended to
the neutrino and electron-neutrino asymmetry coefficients. We
shall not go further into this because it has remained customary
to present experimental results for the old order zero angular
coefficients~\cite{pdg06},

\begin{equation}
B_0 =
\frac{2\lambda(\lambda-1)}{1+3\lambda^2},
\label{bsub0}
\end{equation}

\begin{equation}
a_0 =
\frac{1-\lambda^2}{1+3\lambda^2}.
\label{asub0}
\end{equation}

\noindent
Also, instead of presenting results for $\alpha_e$ it is
customary to give directly the value for $\lambda$, after all
corrections contained in $\alpha_e$ have been applied to the
experimental analysis. Thus, the relevance of exhibiting
Eq.~(\ref{alphae}) is to show that the experimental value of
$\lambda$ is free of theoretical uncertainties at the $10^{-4}$
level.

Another very important constraint for our work here is the
unitarity of the CKM matrix, which we shall use in the form

\begin{equation}
V_{\rm ub} = \sqrt{1 - V^2_{\rm ud} - V^2_{\rm us}}.
\label{Vub}
\end{equation}

\noindent
Given the experimental values of $V_{\rm ub}$ and $V_{\rm us}$,
the only free parameter in Eq.~(\ref{Vub}) is $V_{\rm ud}$.

The current experimental situation~\cite{pdg06} for
Eqs.~(\ref{R}), (\ref{bsub0}), (\ref{asub0}), and (\ref{Vub}) is
given by $R=1.12905(102)\times 10^{-3}\,{\rm s}^{-1}$,
$B_0=0.981(4)$, $a_0=-0.103(4)$, $V_{\rm ub}=0.00431(30)$, and
$V_{\rm us}=0.2257(21)$. It is this last number that recently
increased by three standard deviations from its previous value
and whose effect on the SMR we are going to determine. The
experimental situation of $\lambda$ is at present ambiguous. Its
four more precise determinations are $\lambda_{\rm
A}=-1.2739(19)$~\cite{abele02}, $\lambda_{\rm
L}=-1.266(4)$~\cite{liaud97}, $\lambda_{\rm
Y}=-1.2594(38)$~\cite{yerozolimsky97}, and $\lambda_{\rm
B}=-1.262(5)$~\cite{bopp86}. The last three are statistically
compatible and produce an average $\lambda_{\rm
LYB}=-1.2624(24)$. This average is not statistically compatible
with the value $\lambda_{\rm A}$. Although one may quote an
average of the four $\lambda_{\rm ALYB}=-1.2695(15)$, one must
remember that such an average is not a consistent one. Even so,
it will still be interesting to discuss it.

To determine the SMR we shall form a $\chi^2$ function with the
six constraints Eqs.~(\ref{R}), (\ref{bsub0}), (\ref{asub0}),
(\ref{Vub}), $V^{\rm exp}_{\rm ub}$, and $V^{\rm exp}_{\rm us}$.
This is an over constrained system of restrictions for three
free parameters $\lambda$, $V_{\rm ud}$, and $V_{\rm us}$. This
function is

\begin{eqnarray}
\chi^2 &=& \left(\frac{R'-R}{\sigma_{R'}}\right)^2 +
\left(\frac{\lambda'-\lambda}{\sigma_{\lambda'}}\right)^2 +
\left(\frac{B^{\rm exp}_0-B_0}{\sigma_{B_0}}\right)^2 +
\left(\frac{a^{\rm exp}_0-a_0}{\sigma_{a_0}}\right)^2 +
\nonumber
\\
& &
+ \left(\frac{V^{\rm exp}_{\rm ub}-V_{\rm ub}}
{\sigma_{V_{\rm ub}}}\right)^2 +
\left(\frac{V^{\rm exp}_{\rm us}-V_{\rm us}}
{\sigma_{V_{\rm us}}}\right)^2.
\label{chi2}
\end{eqnarray}

\noindent
The SMR is determined by minimizing ${\chi}^{2}$ at a fine
lattice of points ($\lambda'$, $R'$) in the ($\lambda$, $R$)
plane. It will correspond to the $90\%$ CL region in this plane.
That is, within this region the SM may
be expected to remain valid at the $90\%$ CL.

The key element in the determination of the SMR is that
$\sigma_{R'}$ and $\sigma_{\lambda'}$ are not limited to take
their current experimental values $\sigma_{R}=0.00102\times
10^{-3}\,{\rm s}^{-1}$ and $\sigma_{\lambda}$ around
0.0024. We are at liberty to reduce them down to their
theoretical uncertainty, namely, a few parts at $10^{-4}$. The
theoretically predicted SMR will correspond to $\sigma_{R'}$ and
$\sigma_{\lambda'}$ at approximately one-tenth of their current
experimental counterparts.

In the next two sections we shall study the effects of $V_{\rm
us}$ on the determination of the SMR. Its central value will
affect its position in the ($\lambda$, $R$) plane and its error
bar will affect its width.

\section{\label{secthree}$\bm{V_{\rm us}}$ and the standard model
region}

We shall work within a rectangle of the ($\lambda$, $R$) plane.
The side for $\lambda$ will be $(-1.2744, -1.2552)$ due to the
ambiguity of the experimental value of $\lambda$~\cite{range}. We
shall fold by quadratures the theoretical uncertainty of $R$ into
its experimental error bar to get an effective
$\sigma_{R}=0.00132\times 10^{-3}\,{\rm s}^{-1}$. The other side
of the rectangle will cover three effective standard deviations
above and below the central value of $R=1.12905\times
10^{-3}\,{\rm s}^{-1}$.

To study the effect of $V_{\rm us}$ upon the SMR we shall let its
central value vary up to three standard deviations
$\sigma_{V_{\rm us}}=0.0021$ above and below its current central
value $V_{\rm us}=0.2257$. The other three restrictions
in~(\ref{chi2}) will be kept fixed at their current experimental
values.

There is no need to present all the details of our numerical
analysis. Our results are well illustrated by exhibiting three
cases for the central value of $V_{\rm us}$, namely, 0.2194,
0.2257, and 0.2320 (the first one corresponds to the previous
value of $V_{\rm us}$, the second one to its current value, and
the third one allows for still another three-sigma increase of
$V_{\rm us}$). In each of these cases we use the liberty we have
to choose the size of $\sigma_{R'}$ and $\sigma_{\lambda'}$. The
first choice for them is the corresponding experimental error
bars 0.00132 and 0.0024. The resulting SMR could
well be referred to as the \lq\lq experimental" SMR. The second
choice is to use one-tenth of these values, which as discussed in
the last section is the theoretical SMR. And for the purpose of
further discussion we use as a third choice one-hundredth of
such values.

\begin{figure}
\centerline{\psfig{file=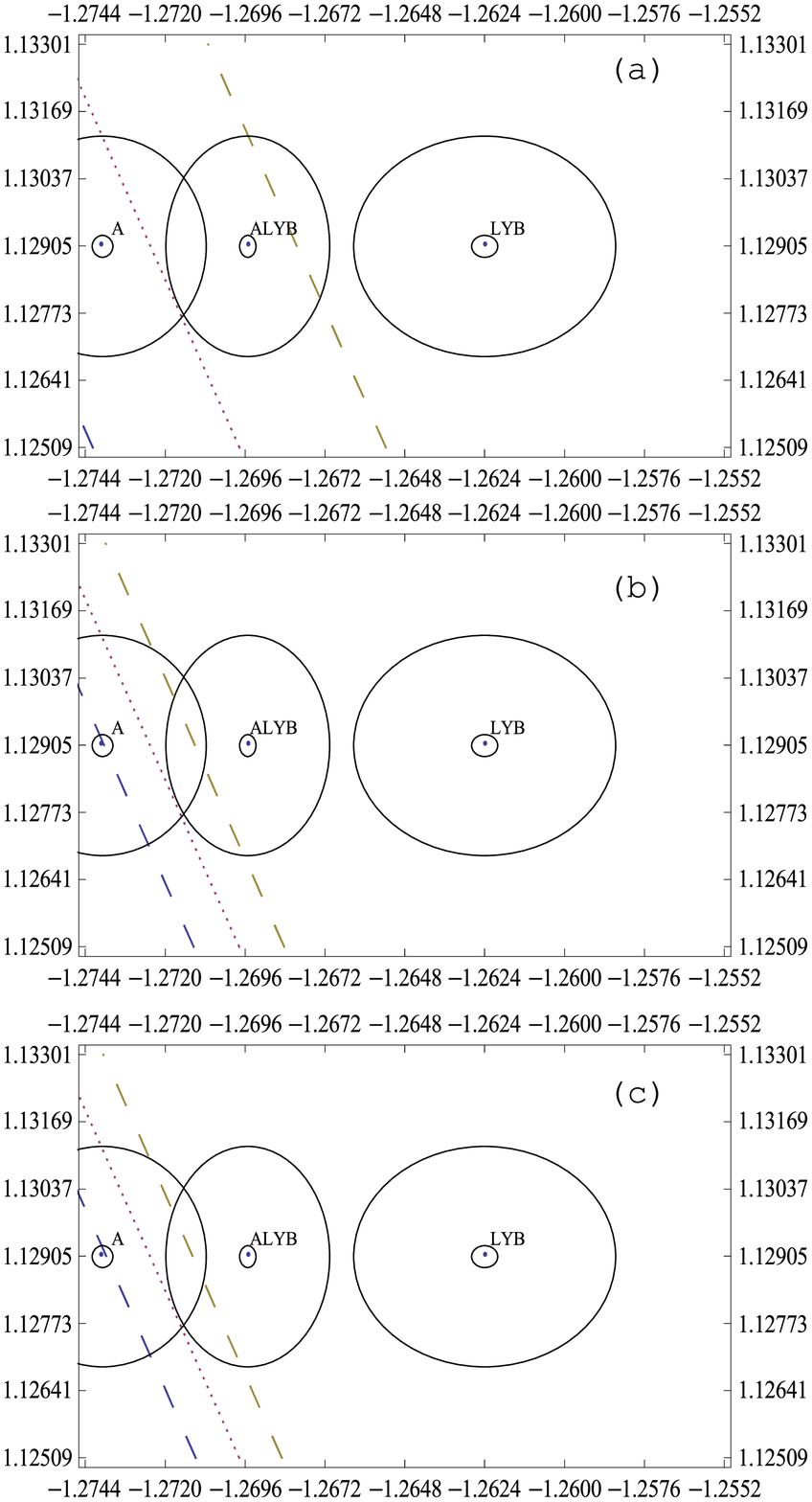,width=6.4in}}
\caption{\label{fig_7}
The detailed numerical results corresponding to
Table~\ref{tabla17} are plotted here. The upper, middle, and
lower entries correspond to (a), (b), and (c), respectively.}
\end{figure}

Our numerical results are given in Tables~\ref{tabla11},
\ref{tabla14}, and \ref{tabla17}. The rows correspond to steps of
one standard deviation in $R'$, $\lambda_0$ gives the
corresponding position of the minimum $\chi^2_0$, the $90\%$ CL
ranges of $\lambda'$ are given in the column headed by $\lambda'$.
In each case the SMR is a band. This can be visualized in the
corresponding Figures~\ref{fig_1}, \ref{fig_4}, and \ref{fig_7}.

To appreciate the variation of $\chi^2$ within the rectangle in
the ($\lambda$, $R$) plane we list its value at sample points in
Tables~\ref{tabla2}, \ref{tabla24}, and \ref{tabla27}. In these
tables one can see how the SMR is narrowed as $\sigma_{R'}$ and
$\sigma_{\lambda'}$ are reduced from their experimental values to
one-tenth of them. But, one also sees that reducing them further
produces no significant narrowing any more.

For comparison purposes, we include in Figures~\ref{fig_1},
\ref{fig_4}, and \ref{fig_7} the $90\%$ CL region around the
central values of the current measurements and, also, the same
regions at one-tenth of the present error bars. Although the
effect of changing $V_{\rm us}$ is perceptible for the \lq\lq
experimental" SMR in Figs.~\ref{fig_1}~(a), \ref{fig_4}~(a), and
\ref{fig_7}~(a) it does not lead to sharp conclusions, unless
$V_{\rm us}$ were to reach $0.2320$. In contrast, the theoretical
SMR of Figs.~\ref{fig_1}~(b), \ref{fig_4}~(b), and
\ref{fig_7}~(b) clearly discriminate $\lambda_{\rm A}$ and
$\lambda_{\rm LYB}$. The current situation is depicted in
Fig.~\ref{fig_4}~(b). $\lambda_{\rm LYB}$ is sharply incompatible
with the SM. Thus, either the SM is quite accurate and
$\lambda_{\rm LYB}$ will be eliminated or, if this $\lambda_{\rm
LYB}$ is confirmed in the future, n$\beta$d will produce strong
evidence for not too far away new physics. Correspondingly, if
$\lambda_{\rm A}$ is confirmed in the future, the accuracy of the
SM will be sustained and new physics will be farther away; so it
will be harder to detect it in n$\beta$d. The above disjunctive
is further strengthened if $V_{\rm us}$ is measured still higher,
as seen in Fig.~\ref{fig_7}~(b). Notice that the current central
values of $\lambda_{\rm A}$ and $\lambda_{\rm LYB}$, if either of
them were to be confirmed, strongly indicate the existence of new
physics, as can be appreciated with the small regions around them
in Fig.~\ref{fig_4}~(b). The SM would remain very accurate if
$\lambda_{\rm A}$ were confirmed and $V_{\rm us}$ were further
increased up to $0.2320$. This possibility is illustrated in
Fig.~\ref{fig_7}~(c). Surprisingly, the inconsistent average
$\lambda_{\rm ALYB}$ is fully compatible with the SM at present,
as seen in Fig.~\ref{fig_4}~(b).

That arbitrarily reducing $\sigma_{R'}$ and $\sigma_{\lambda'}$
up to one-hundredth of their experimental counterparts produced
no significant reduction of the SMR, as can be seen in
Figs.~\ref{fig_1}~(c), \ref{fig_4}~(c), and \ref{fig_7}~(c),
requires some detailed discussion. The reason for this can be
traced to the individual contributions of $R'$, $\lambda'$, and
$V_{\rm us}$ to $\chi^2$ of Eq.~(\ref{chi2}). In this respect, we
have produced Table~\ref{tablavii}. It is sufficient to present
the case of the central row in Table~\ref{tabla14}, where
$V_{\rm us} = V^{\rm exp}_{\rm us} = 0.2257$ and $R = R^{\rm
exp}=1.12905\times 10^{-3}\,{\rm s}^{-1}$, and the
contributions to $\chi^2$ at the border of the SMR, namely, the
extremes of the corresponding ranges of $\lambda'$ in
Table~\ref{tabla14}.

In the top part of Table~\ref{tablavii} we give the six different
contributions to $\chi^2$ at the above extremes. One can see that
with $\sigma_{R'}$ and $\sigma_{\lambda'}$ at their experimental
values the $\chi^2(R')$ and $\chi^2(\lambda')$ contributions
dominate over the $\chi^2(V_{\rm us})$ contribution (upper
entries). At $1/10$ of these values the situation is reversed and
it remains so when $\sigma_{R'}$ and $\sigma_{\lambda'}$ are
reduced up to $1/100$ (second and third entries). In the lower
part of Table~\ref{tablavii} we trace in more detail when this
reversal takes place by reducing $\sigma_{R'}$ and
$\sigma_{\lambda'}$ by $1/2$, $1/5$, and $1/7$ (second, third,
and fourth entries, respectively). The dominance of
$\chi^2(V_{\rm us})$ over $\chi^2(R')$ and $\chi^2(\lambda')$
takes place already when $\sigma_{R'}$ and $\sigma_{\lambda'}$
are cut to between $1/4$ and $1/5$ of their experimental
counterparts. Notice that this reversal does not depend on $B_0$,
$a_0$, and $V_{\rm ub}$, whose $\chi^2$ contributions remain
fairly constant throughout Table~\ref{tablavii}. One may conclude
that the potential of the SM prediction at $1/10$ of the
experimental errors on $R$ and $\lambda$ cannot be reached,
because of the current uncertainty on $V_{\rm us}$. In other
words, even if the experimental precision in n$\beta$d were to be
greatly improved in the near future, the comparison with the SM
predictions will be severely limited by the experimental
precision of $V_{\rm us}$.

Let us next study in detail the effects of improving the
precision of $V_{\rm us}$.

\section{\label{secfour}The precision of $\bm{V_{\rm us}}$ and the
standard model region}

n$\beta$d cannot provide a better test of the SM even if the
error bars on $R$ and $\lambda$ and the theoretical uncertainty
in Eq.~(\ref{R}) were to be reduced beyond one-fifth. As seen in
the previous section, the limitation comes from the error bars on
$V_{\rm us}$. The central value of $V_{\rm us}$ does shift the
position of the SMR, but it is reducing $\sigma_{V_{\rm us}}$
that will improve the width of the SMR.

To see this we have reproduced the SMR assuming $\sigma_{V_{\rm
us}}$ is cut to one-tenth of its current value, that is
$\sigma_{V_{\rm us}}=0.00021$, and assuming the central value to
be at three places, $V_{\rm us}=0.2194$, 0.2257, or 0.2320. Of
course this last is only an assumption, all we can say as of now
is that such central value will fall at $90\%$ CL somewhere within
the band of Fig.~\ref{fig_4}~(b). The corresponding numerical
results are summarized in Tables~\ref{tabla11v9},
\ref{tabla14v9}, and \ref{tabla17v9} for $\chi^2_0$, $\lambda_0$,
and the $90\%$ CL
range of $\lambda'$. Values of $\chi^2$ at sample points in the
($\lambda$, $R$) plane are found in Tables~\ref{tabla2v9},
\ref{tabla24v9}, and \ref{tabla27v9}. In each row of these six
tables the upper, middle, and lower entries correspond to
$\sigma_{R'}$ and $\sigma_{\lambda'}$ at $\sigma_{R}$ and
$\sigma_{\lambda}$, at $\sigma_{R}/10$ and
$\sigma_{\lambda}/10$, and at $\sigma_{R}/100$ and
$\sigma_{\lambda}/100$, respectively. Notice that the numerical
values of $\chi^2_0$ and $\lambda_0$ are practically the same in
Tables~\ref{tabla11} and \ref{tabla11v9},
\ref{tabla14} and \ref{tabla14v9}, and \ref{tabla17} and
\ref{tabla17v9}. The minimum of $\chi^2$ and the position of the
minimum in the ($\lambda$, $R$) plane are practically independent
of the values of $\sigma_{R'}$, $\sigma_{\lambda'}$, and
$\sigma_{V_{\rm us}}$. In contrast, the values of $\chi^2$ at
sample points in the ($\lambda$, $R$) plane away from the SMR
become enormous, as can be appreciated looking throughout
Tables~\ref{tabla2v9}, \ref{tabla24v9}, and \ref{tabla27v9}. Such
increases in $\chi^2$ indicate the substantial narrowing of the
SMR as $\sigma_{V_{\rm us}}$ is reduced along with $\sigma_{R'}$
and $\sigma_{\lambda'}$. These results can be visualized
in Figs.~\ref{fig_1v9}, \ref{fig_4v9}, and \ref{fig_7v9}.
Comparing these last figures with the corresponding ones of
Sec.~\ref{secthree}, one sees that the \lq\lq experimental" SMR
is not noticeably reduced, as was to be expected. However, at
one-tenth $\sigma_{R'}$ and $\sigma_{\lambda'}$, the comparison
of Figs.~\ref{fig_1}~(b), \ref{fig_4}~(b), and \ref{fig_7}~(b),
with Figs.~\ref{fig_1v9}~(b), \ref{fig_4v9}~(b), and
\ref{fig_7v9}~(b), respectively, shows that the effect of
reducing $\sigma_{V_{\rm us}}$ is quite impressive. As seen
in Tables~\ref{tabla2v9}, \ref{tabla24v9}, and \ref{tabla27v9},
the SMR is greatly reduced. This reduction of the SMR could lead
to almost a thin line if the theoretical and experimental
uncertainties in $R$ and $\lambda$ were put under much better
control, as can be visualized in Figs.~\ref{fig_1v9}~(c),
\ref{fig_4v9}~(c), and \ref{fig_7v9}~(c).

There is a systematic feature in
Tables~\ref{tabla11}-\ref{tabla17} and
Tables~\ref{tabla11v9}-\ref{tabla17v9}, the value of $\chi^2_0$
is always around $2.90$. The reason for this is found in
Table~\ref{tablavii}, the contribution of the neutrino asymmetry
$B_0$ to $\chi^2$ is always around $2.70$. This is a $1.6$
standard deviations from the SM prediction. It is not significant
and we shall not discuss it further.

It is clear that the ability of n$\beta$d to test the SM is
intimately connected with the precision to determine $V_{\rm us}$
in strangeness-changing decays.

\section{\label{secfive}Predictions of $\bm{V_{\rm us}}$ from
neutron beta decay}

A precise determination of $V_{\rm us}$ in strangeness-changing
decays may take longer than precise measurements of $R$ and
$\alpha_e$ or $\lambda$. n$\beta$d may provide a better
determination of $V_{\rm us}$ via the unitarity of
the CKM matrix, once the former produce a precise measurement of
$V_{\rm ud}$. This is a complementary way to appreciate the
results of the last two sections.

First, let us look into the current determination of $V_{\rm
ud}$. The ambiguity in $\lambda$ leads to an ambiguity in the
experimental value of $V_{\rm ud}$. One has correspondingly two
incompatible values for $V_{\rm ud}$, namely,

\begin{equation}
V^{\rm LYB}_{\rm ud} = 0.9791\pm 0.0016
\label{Vudlyb}
\end{equation}

\noindent
and

\begin{equation}
V^{\rm A}_{\rm ud} = 0.9718\pm 0.0013.
\label{Vuda}
\end{equation}

\noindent
One may also quote the third, albeit inconsistent, value

\begin{equation}
V^{\rm ALYB}_{\rm ud} = 0.9746\pm 0.0011.
\label{Vudalyb}
\end{equation}

\noindent
Although, not yet satisfactory, one can already see that the
error bars are competitive with $V_{\rm ud}$ determined from
other sources~\cite{pdg06}. Also, within the validity of the SM,
these values are accompanied by

\begin{equation}
V^{\rm LYB}_{\rm us} = 0.2032\pm 0.0079,
\label{Vuslyb}
\end{equation}

\begin{equation}
V^{\rm A}_{\rm us} = 0.2357\pm 0.0055,
\label{Vusa}
\end{equation}

\noindent
and

\begin{equation}
V^{\rm ALYB}_{\rm us} = 0.2239\pm 0.0048.
\label{Vusalyb}
\end{equation}

\noindent
Again, even if not satisfactory, the error bars are becoming
competitive with $V_{\rm us}$ determined from other
sources~\cite{pdg06}.

Let us match Eqs.~(\ref{Vuslyb})-(\ref{Vusalyb}) with the value
of $V_{\rm us}$ from $K_{l3}$ decays (which was used in the
previous sections), namely,

\begin{equation}
V^{K_{l3}}_{\rm us} = 0.2257\pm 0.0021.
\label{Vuskl3}
\end{equation}

\noindent
It is convenient to produce the $90\%$~CL ranges that correspond
to these $V_{\rm us}$ values. They are

\begin{equation}
V^{\rm LYB}_{\rm us}(90\%\, {\rm CL}) = (0.1905,\, 0.2159),
\label{Vuslyb90}
\end{equation}

\begin{equation}
V^{\rm A}_{\rm us}(90\%\, {\rm CL}) = (0.2270,\, 0.2444),
\label{Vusa90}
\end{equation}

\begin{equation}
V^{\rm ALYB}_{\rm us}(90\%\, {\rm CL}) = (0.2163,\, 0.2315),
\label{Vusalyb90}
\end{equation}

\noindent
and

\begin{equation}
V^{K_{l3}}_{\rm us}(90\%\, {\rm CL}) = (0.2222,\, 0.2292).
\label{Vuskl390}
\end{equation}

\noindent
One can readily see that range~(\ref{Vuslyb90}) is below
(\ref{Vuskl390}) and there is no overlap between them at all.
Range~(\ref{Vusa90}) is above (\ref{Vuskl390}) and there is a
small overlap between the two. Contrastingly, range~(\ref{Vusalyb90})
fully contains range (\ref{Vuskl390}). These comparisons
correspond to the overlapping or lack of it of the $90\%$ CL
ellipses with the SMR exhibited in Fig.~\ref{fig_4}~(b).

Also, they indirectly exhibit the current experimental problem in
the determination of $\lambda$. Ranges (\ref{Vuslyb90}) and
(\ref{Vusa90}) do not overlap with one another and are quite
separated. These comparisons are complementary to the analysis of
sections \ref{secthree} and \ref{secfour}. They provide a quick
way to see the compatibility of n$\beta$d data together with
$K_{l3}$ data with the SM assumptions.

\begin{figure}
\centerline{\psfig{file=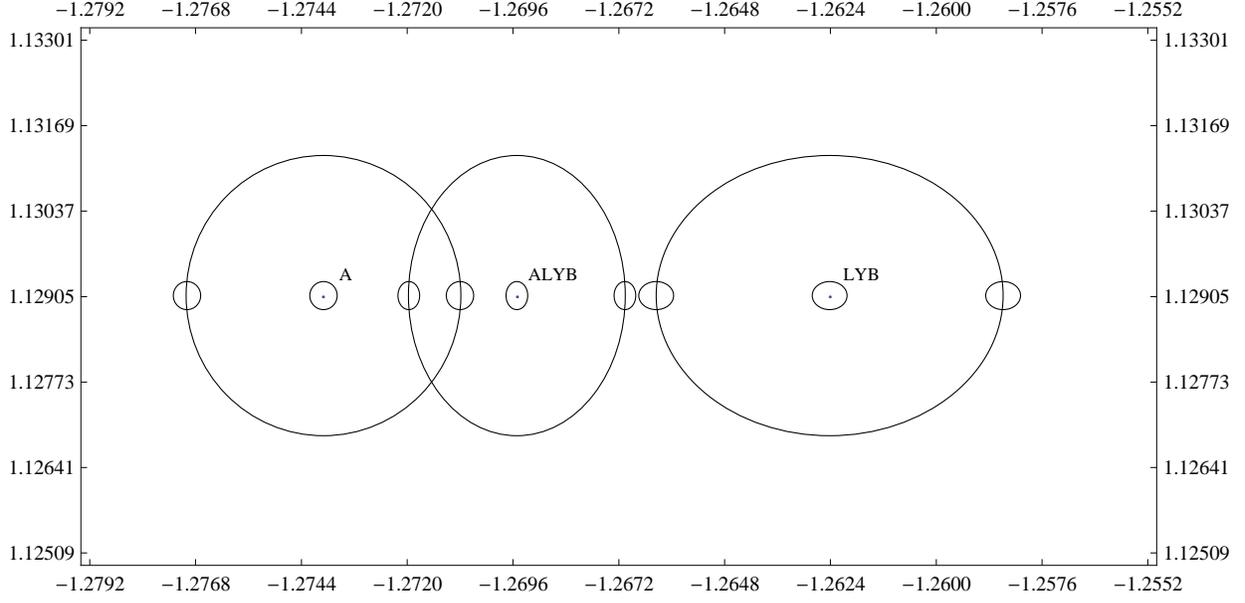,width=6.4in}}
\caption{\label{fig_num7}
Current $90\%$~CL regions around $\lambda_{\rm LYB}$,
$\lambda_{\rm A}$, and $\lambda_{\rm ALYB}$ and $90\%$~CL regions
around the central and horizontal border points when
$\sigma_{R}$ and $\sigma_ {\lambda}$ are cut to $1/10$ of their
present values.}
\end{figure}

The present experimental situation will be corrected eventually.
In the meantime, we can extend this analysis through $V_{\rm
us}$. To appreciate what can be expected we have produced a set
of values for $V_{\rm ud}$ and $V_{\rm us}$ assuming the central
values of $R$ and $\lambda$ are at the left- and right-hand and
at the center of the $90\%$ CL
ranges of $\lambda_{\rm LYB}$, $\lambda_{\rm A}$, and
$\lambda_{\rm ALYB}$. The former two are indicated by a $-$ and a
$+$ sign, respectively. The corresponding error bars are
$\sigma_{R}/10$ and $\sigma_{\lambda}/10$. These points and their
$90\%$ CL regions
are displayed in Fig.~\ref{fig_num7}. The numerical results are
exhibited in Table~\ref{tablaxiv}.

The main result that can be seen in this table is the size of the
error bars of $V_{\rm ud}$ and $V_{\rm us}$. $\sigma_{V_{\rm
ud}}$ is reduced to around $0.0002$, which is between $1/5$ and
$1/6$ of the error bars of Eqs.~(\ref{Vudlyb})-(\ref{Vudalyb}).
$\sigma_{V_{\rm us}}$ is reduced to around $0.0008$, which is
between $1/2$ and $1/3$ of the current error bar of $0.0021$ of
Eq.~(\ref{Vuskl3}). Clearly, once n$\beta$d produces a
consistent value for $V_{\rm ud}$ its potential precision will
improve substantially over its determination from other sources.
Assuming CKM-matrix unitarity, its accompanying value for $V_{\rm
us}$ will improve over its current determination from
strangeness-changing decays and may remain so for sometime. This
value will be useful in calculations that assume the validity of
the SM and in coming tests of the unitarity triangle. A direct
comparison with the independently improved future determinations
of $V_{\rm us}$ from strangeness-changing decays will readily
indicate if signals of new physics are present or not.

\section{\label{summary}Summary and discussion}

n$\beta$d data and $K_{l3}$ data are two sets of independent data
and each one by itself cannot test the SM. So, it is not a
question of whether the former is compatible with the latter.
Only using the two sets simultaneously can provide tests on the
SM and the question is if their simultaneous use is compatible
with the SM assumptions. Such compatibility can be fully seen
through the overlap of the $90\%$ CL ellipses around precise
experimental determinations of $R$ and $\lambda$ with the band of
the SMR, which requires precise $V_{\rm us}$ determinations in
strangeness-changing decays and in particular in $K_{l3}$ decays.
The non-overlapping of these two regions would give signals of
physics beyond the SM.

\begin{figure}
\centerline{\psfig{file=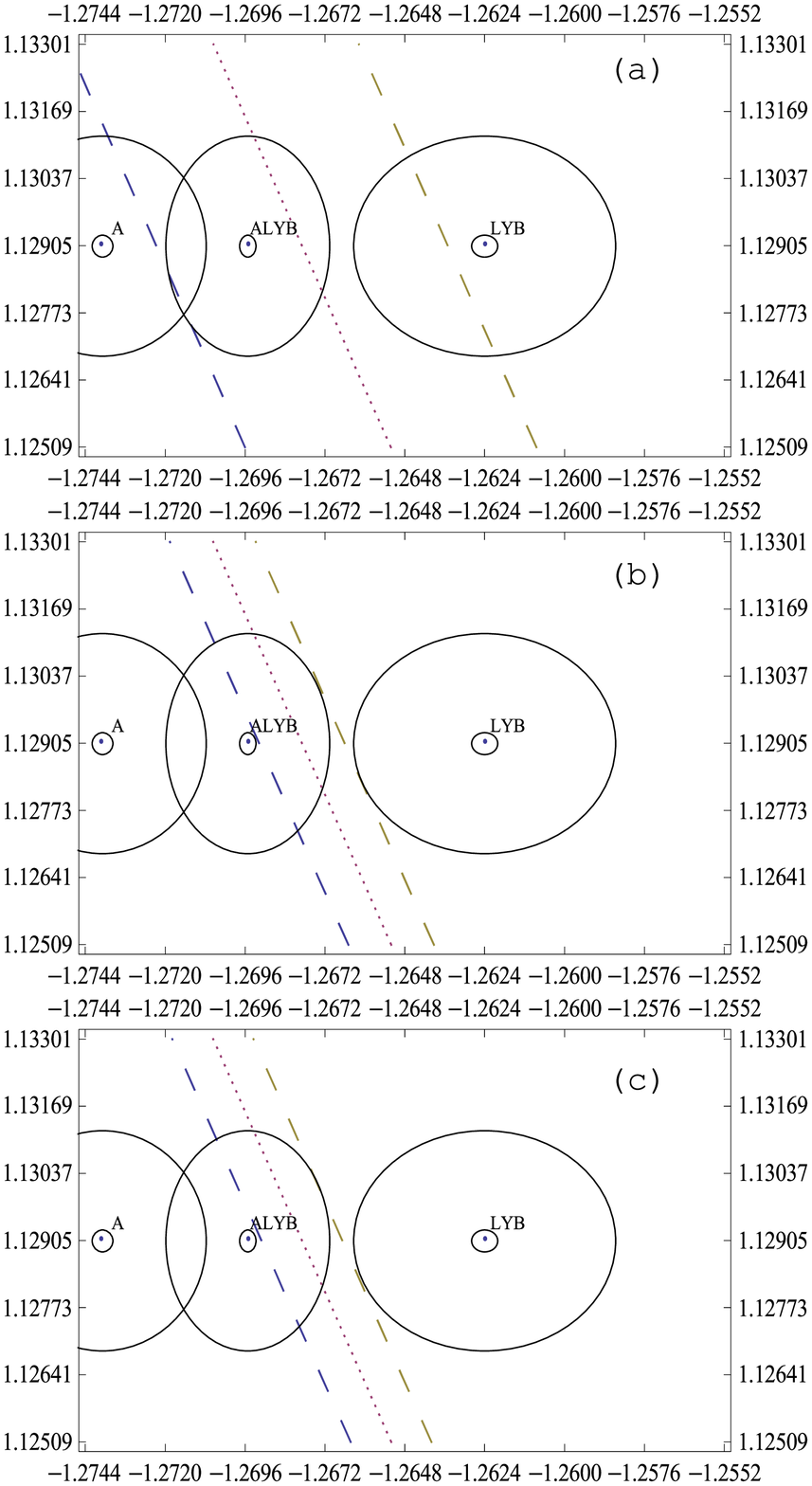,width=6.4in}}
\caption{\label{fig_1}
The detailed numerical results corresponding to
Table~\ref{tabla11} are plotted here. The upper, middle, and
lower entries correspond to (a), (b), and (c), respectively.}
\end{figure}

\begin{figure}
\centerline{\psfig{file=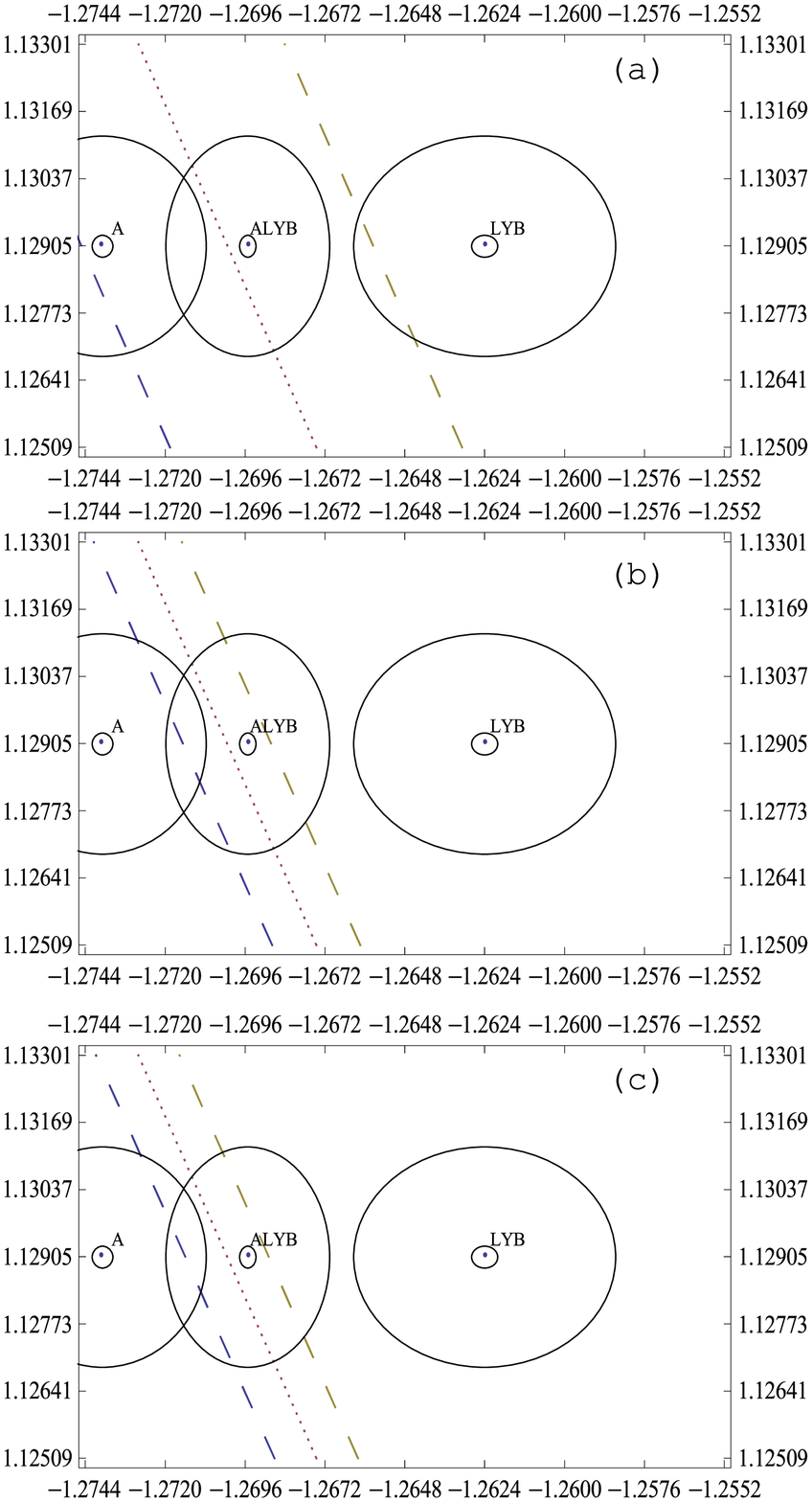,width=6.4in}}
\caption{\label{fig_4}
The detailed numerical results corresponding to
Table~\ref{tabla14} are plotted here. The upper, middle, and
lower entries correspond to (a), (b), and (c), respectively.}
\end{figure}

The current potential of n$\beta$d to discover new physics is
seen in the overlap of the $90\%$ CL regions around
$\lambda_{\rm A}$ and $\lambda_{\rm LBY}$ with the theoretical
SMR in Fig.~\ref{fig_4}(b). The recent change of three standard
deviations in $V_{\rm us}$ can be appreciated in the shift of
the SMR from Fig.~\ref{fig_1}(b) to Fig.~\ref{fig_4}(b). This
shift is towards $\lambda_{\rm A}$, meaning that $\lambda_{\rm
LBY}$ is either ruled out by the accuracy of the SM or it gives
a strong signal for new physics. In contrast, $\lambda_{\rm A}$
favors such an accuracy and, if confirmed in the future, it means
that new physics is farther away.

However, the current potential is limited by the experimental
precision of $V_{\rm us}$. Actually, if such precision is not
improved, reducing the error bars on $R$ and $\lambda$ beyond
$1/4$ or $1/5$ of their current values will not lead to better
tests of the SM. However, if this precision is improved in the
future to somewhere between $1/2$ and $1/3$ of what it is at
present, then n$\beta$d will provide tests of the SM at the level
of the value of $V_{\rm us}$ it can produce, via CKM-matrix
unitarity, as can be appreciated from the combined analysis of
sections~\ref{secthree}-\ref{secfive}.

\begin{figure}
\centerline{\psfig{file=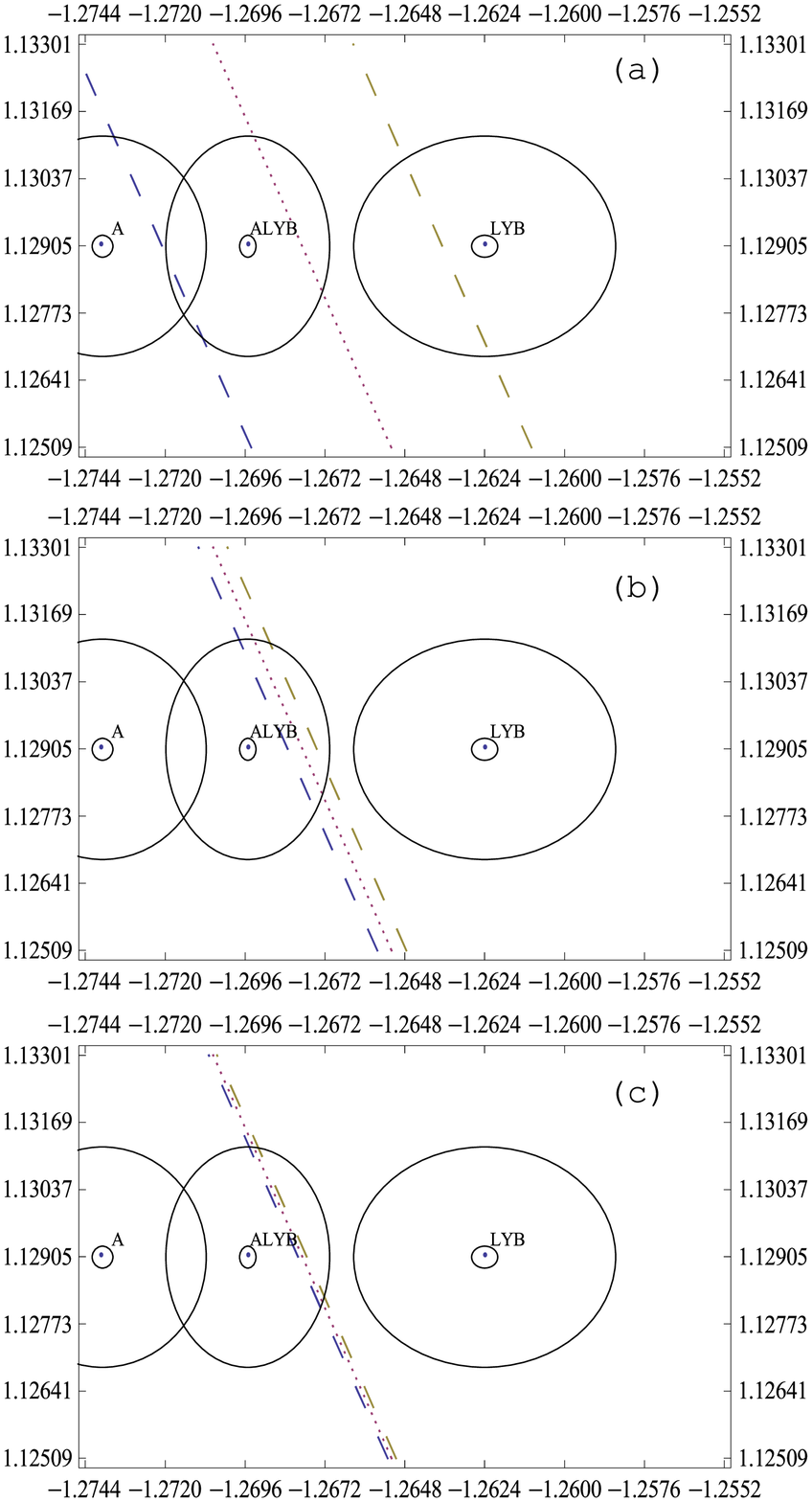,width=6.4in}}
\caption{\label{fig_1v9}
These figures correspond to Figs.~\ref{fig_1}~(a),
\ref{fig_1}~(b), and \ref{fig_1}~(c) when $\sigma_{V_{\rm us}}$
is assumed to be at $0.00021$.} \end{figure}

\begin{figure}
\centerline{\psfig{file=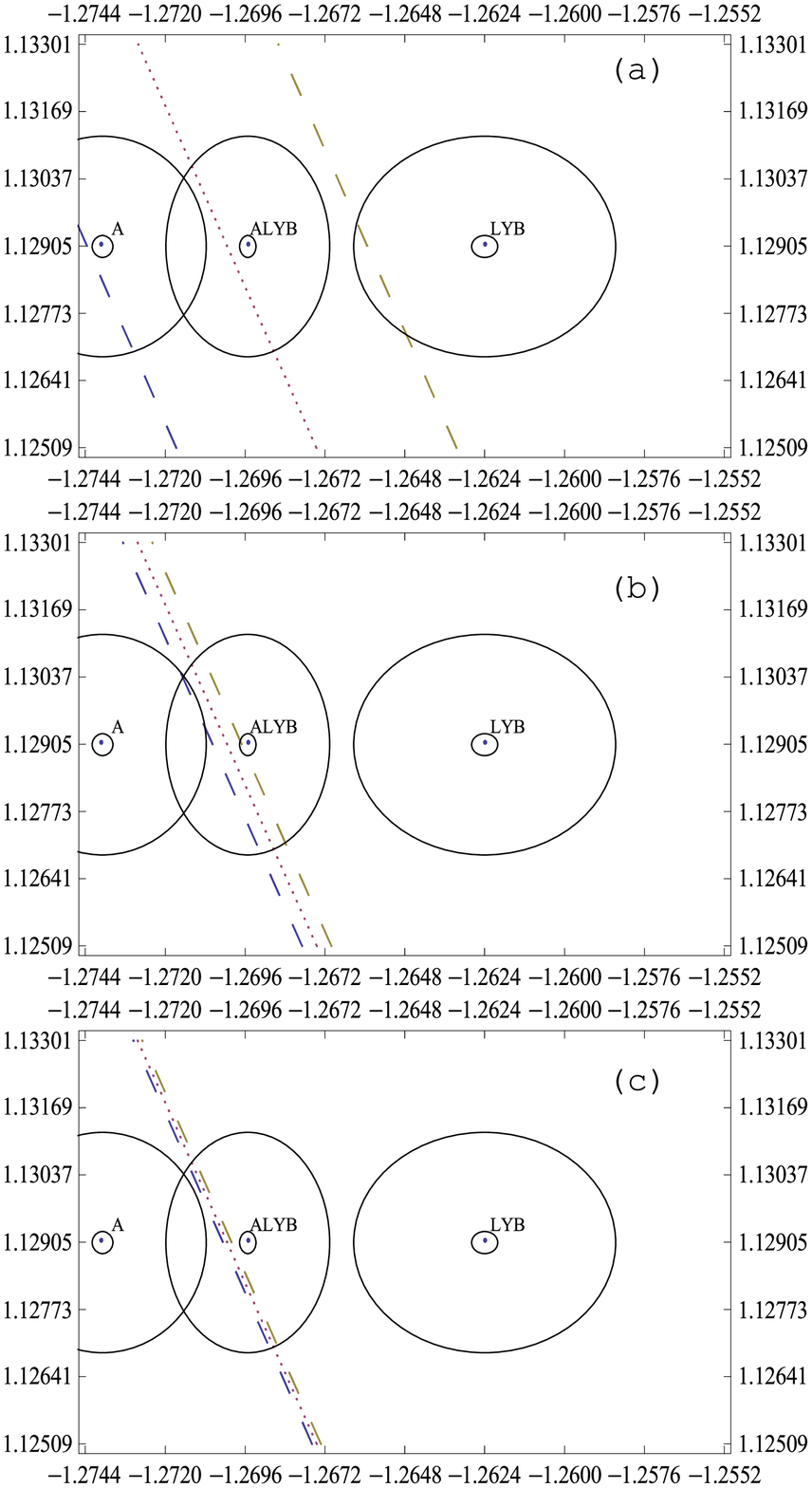,width=6.4in}}
\caption{\label{fig_4v9}
These figures correspond to Figs.~\ref{fig_4}~(a),
\ref{fig_4}~(b), and \ref{fig_4}~(c) when $\sigma_{V_{\rm us}}$
is assumed to be at $0.00021$.} \end{figure}

\begin{figure}
\centerline{\psfig{file=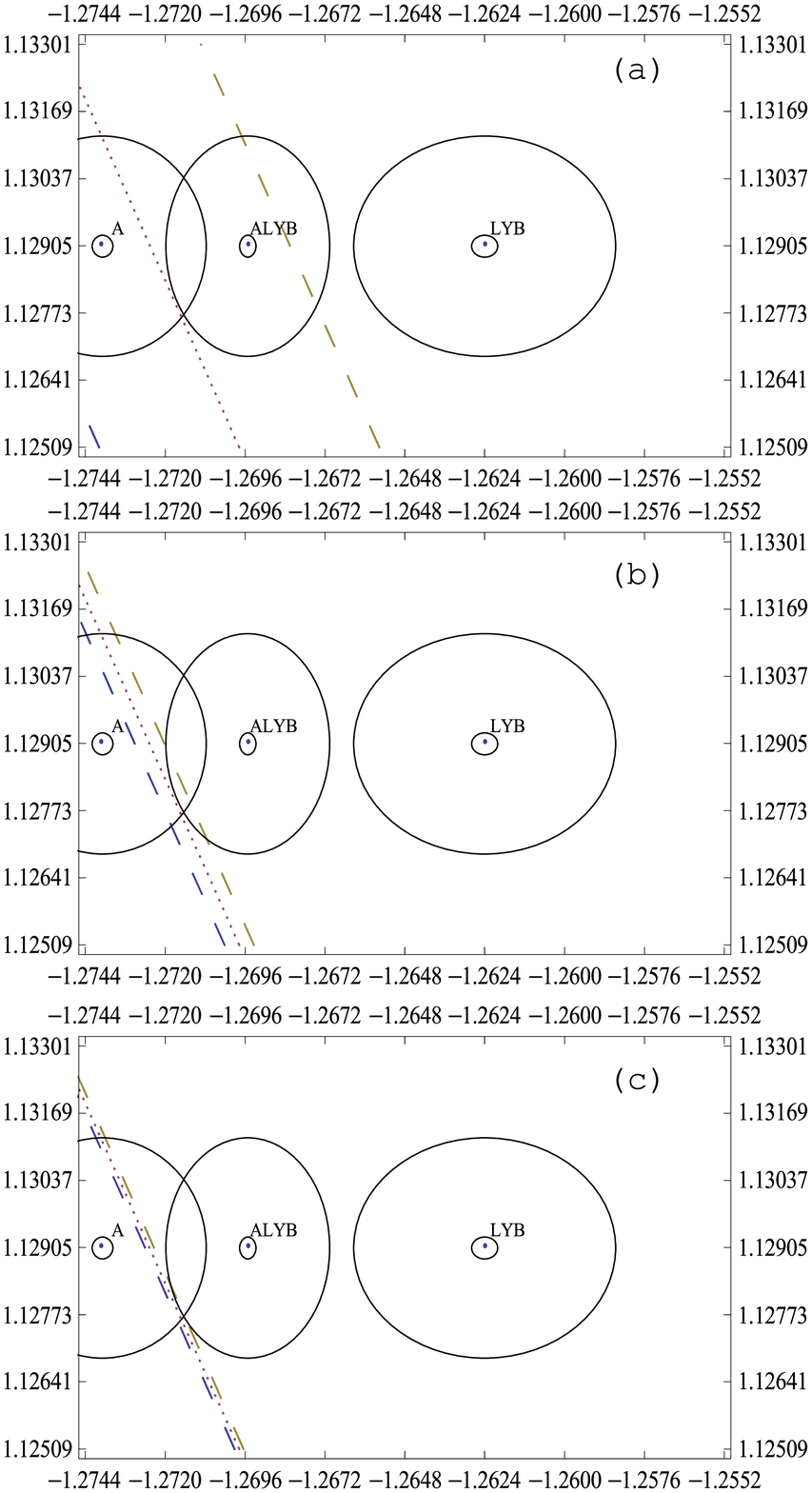,width=6.4in}}
\caption{\label{fig_7v9}
These figures correspond to Figs.~\ref{fig_7}~(a),
\ref{fig_7}~(b), and \ref{fig_7}~(c) when $\sigma_{V_{\rm us}}$
is assumed to be at $0.00021$.} \end{figure}

The full potential of the SMR to confirm the accuracy of the SM
is seen when $\sigma_{V_{\rm us}}$ is reduced further. If
eventually strangeness-changing decays are to reduce
$\sigma_{V_{\rm us}}$ to $1/10$ of its current value, then the
SMR becomes a very thin band. This can be visualized in
Figs.~\ref{fig_1v9}-\ref{fig_7v9}. When this occurs, n$\beta$d
combined with strangeness-changing decays will provide very
severe tests of the SM and may detect new physics which for
whatever reason is very far away.

Before the above situation occurs, n$\beta$d may produce a
prediction for $V_{\rm us}$ via the unitarity of the CKM-matrix.
Such a prediction may be useful, while the experimental $V_{\rm
us}$ remains at its current value, in calculations that assume
the validity of the SM and in other tests of the SM through the
unitarity triangle. Also, even if n$\beta$d data are independent
of $K_{l3}$ data, this prediction of $V_{\rm us}$ with n$\beta$d
data may appear to be incompatible with the measurement of $V_{\rm us}$
in $K_{l3}$. This apparent incompatibility of n$\beta$d and
$K_{l3}$ decays would provide a quick indication of the necessity
to go beyond the SM.

Even if the present situation in n$\beta$d is not satisfactory,
ideally, in the future the combined effort of reducing the
theoretical and experimental error will produce a SMR close to a
line, as can be seen in
Figs.~\ref{fig_1v9}~(c)-\ref{fig_7v9}~(c). Difficult as this
task may seem, it does show the potential low energy physics has
to test the SM.

\begin{acknowledgments}
One of us (G.~S-C) is grateful to the Faculty of Mathematics,
Autonomous University of Yucat\'an, M\'exico, for hospitality
where part of this work was done. A.G.\ and G.~S-C would like to
thank CONACyT (M\'exico) for partial support.
\end{acknowledgments}

\clearpage

{\squeezetable

\begin{table}

\caption{The minimum of $\chi^2$ ($\chi^2_0$) and its
corresponding value of $\lambda$ ($\lambda_0$) for six values of
$R$ (which change in steps of one $\sigma_{R}$). In each row the
upper, middle, and lower entries correspond to the size of error
bars of $R$ and $\lambda$ discussed in the text. The
$90\%$\,CL ranges for $\lambda$ are displayed in the last column.
$V_{\rm us}$ is assumed to be at $V^{\rm exp}_{\rm
us} - 3\,\sigma_{V_{\rm us}}$, with $\sigma_{V_{\rm us}} =
0.0021$.
\label{tabla11}}

\begin{ruledtabular}

\begin{tabular}{cccc}

$R'$ & $\chi^2_0$ & $\lambda_0$ & $\lambda'$ \\

\hline

1.12509 & 2.97456 & $-$1.26522 & ($-$1.26961, $-$1.26085) \\

& 2.97495 & $-$1.26521 & ($-$1.26650, $-$1.26393)  \\

& 2.97496 & $-$1.26521 & ($-$1.26643, $-$1.26400) \\

\\

1.12641 & 2.93935 & $-$1.26612 & ($-$1.27049, $-$1.26174) \\

& 2.93958 & $-$1.26610 & ($-$1.26740, $-$1.26483) \\

& 2.93958 & $-$1.26610 & ($-$1.26733, $-$1.26490) \\

\\

1.12773 & 2.91357 & $-$1.26700 & ($-$1.27138, $-$1.26263) \\

& 2.91368 & $-$1.26699 & ($-$1.26829, $-$1.26572) \\

& 2.91368 & $-$1.26699 & ($-$1.26822, $-$1.26579) \\

\\

1.12905 & 2.89715 & $-$1.26789 & ($-$1.27227, $-$1.26351) \\

& 2.89719 & $-$1.26789 & ($-$1.26918, $-$1.26661) \\

& 2.89719 & $-$1.26789 & ($-$1.26911, $-$1.26668) \\

\\

1.13037 & 2.89005 & $-$1.26878 & ($-$1.27316, $-$1.26440) \\

& 2.89005 & $-$1.26878 & ($-$1.27008, $-$1.26750) \\

& 2.89005 & $-$1.26878 & ($-$1.27001, $-$1.26757) \\

\\

1.13169 & 2.89220 & $-$1.26967 & ($-$1.27405, $-$1.26529) \\

& 2.89221 & $-$1.26967 & ($-$1.27097, $-$1.26839) \\

& 2.89221 & $-$1.26967 & ($-$1.27090, $-$1.26846) \\

\\

1.13301 & 2.90353 & $-$1.27055 & ($-$1.27493, $-$1.26618) \\

& 2.90360 & $-$1.27056 & ($-$1.27186, $-$1.26928) \\

& 2.90360 & $-$1.27056 & ($-$1.27179, $-$1.26935) \\
 
\end{tabular}

\end{ruledtabular}

\end{table}}

\clearpage

{\squeezetable

\begin{table}

\caption{The minimum of $\chi^2$ ($\chi^2_0$) and its
corresponding value of $\lambda$ ($\lambda_0$) for six values of
$R$ (which change in steps of one $\sigma_{R}$). In each row the
upper, middle, and lower entries correspond to the size of error
bars of $R$ and $\lambda$ discussed in the text. The
$90\%$\,CL ranges for $\lambda$ are displayed in the last column.
$V_{\rm us}$ is assumed to be at $V^{\rm exp}_{\rm
us}$, with $\sigma_{V_{\rm us}} = 0.0021$.
\label{tabla14}}

\begin{ruledtabular}

\begin{tabular}{cccc}

$R'$ & $\chi^2_0$ & $\lambda_0$ & $\lambda'$ \\

\hline

1.12509 & 2.90389 & $-$1.26746 & ($-$1.27186, $-$1.26308) \\

& 2.90396 & $-$1.26746 & ($-$1.26879, $-$1.26614)  \\

& 2.90396 & $-$1.26746 & ($-$1.26872, $-$1.26621) \\

\\

1.12641 & 2.89229 & $-$1.26835 & ($-$1.27275, $-$1.26396) \\

& 2.89230 & $-$1.26835 & ($-$1.26969, $-$1.26704) \\

& 2.89230 & $-$1.26835 & ($-$1.26962, $-$1.26711) \\

\\

1.12773 & 2.89003 & $-$1.26924 & ($-$1.27364, $-$1.26486) \\

& 2.89003 & $-$1.26925 & ($-$1.27058, $-$1.26793) \\

& 2.89003 & $-$1.26925 & ($-$1.27051, $-$1.26800) \\

\\

1.12905 & 2.89703 & $-$1.27013 & ($-$1.27452, $-$1.26575) \\

& 2.89707 & $-$1.27014 & ($-$1.27147, $-$1.26882) \\

& 2.89707 & $-$1.27014 & ($-$1.27140, $-$1.26889) \\

\\

1.13037 & 2.91325 & $-$1.27102 & ($-$1.27541, $-$1.26663) \\

& 2.91336 & $-$1.27103 & ($-$1.27237, $-$1.26972) \\

& 2.91336 & $-$1.27103 & ($-$1.27230, $-$1.26978) \\

\\

1.13169 & 2.93862 & $-$1.27191 & ($-$1.27630, $-$1.26752) \\

& 2.93884 & $-$1.27192 & ($-$1.27326, $-$1.27061) \\

& 2.93884 & $-$1.27192 & ($-$1.27317, $-$1.27068) \\

\\

1.13301 & 2.97309 & $-$1.27280 & ($-$1.27719, $-$1.26841) \\

& 2.97347 & $-$1.27281 & ($-$1.27415, $-$1.27150) \\

& 2.97347 & $-$1.27281 & ($-$1.27408, $-$1.27157) \\
 
\end{tabular}

\end{ruledtabular}

\end{table}}

\clearpage

{\squeezetable

\begin{table}

\caption{The minimum of $\chi^2$ ($\chi^2_0$) and its
corresponding value of $\lambda$ ($\lambda_0$) for six values of
$R$ (which change in steps of one $\sigma_{R}$). In each row the
upper, middle, and lower entries correspond to the size of error
bars of $R$ and $\lambda$ discussed in the text. The
$90\%$\,CL ranges for $\lambda$ are displayed in the last column.
$V_{\rm us}$ is assumed to be at $V^{\rm exp}_{\rm
us} + 3\,\sigma_{V_{\rm us}}$, with $\sigma_{V_{\rm us}} =
0.0021$.
\label{tabla17}}

\begin{ruledtabular}

\begin{tabular}{cccc}

$R'$ & $\chi^2_0$ & $\lambda_0$ & $\lambda'$ \\

\hline

1.12509 & 2.89310 & $-$1.26977 & ($-$1.27418, $-$1.26537) \\

& 2.89312 & $-$1.26978 & ($-$1.27115, $-$1.26843)  \\

& 2.89312 & $-$1.26978 & ($-$1.27109, $-$1.26850) \\

\\

1.12641 & 2.90567 & $-$1.27067 & ($-$1.27507, $-$1.26627) \\

& 2.90574 & $-$1.27068 & ($-$1.27205, $-$1.26932) \\

& 2.90574 & $-$1.27068 & ($-$1.27198, $-$1.26939) \\

\\

1.12773 & 2.92746 & $-$1.27156 & ($-$1.27596, $-$1.26715) \\

& 2.92764 & $-$1.27157 & ($-$1.27294, $-$1.27022) \\

& 2.92764 & $-$1.27157 & ($-$1.27288, $-$1.27029) \\

\\

1.12905 & 2.95843 & $-$1.27245 & ($-$1.27685, $-$1.26805) \\

& 2.95875 & $-$1.27246 & ($-$1.27384, $-$1.27111) \\

& 2.95875 & $-$1.27246 & ($-$1.27377, $-$1.27118) \\

\\

1.13037 & 2.99852 & $-$1.27334 & ($-$1.27774, $-$1.26894) \\

& 2.99901 & $-$1.27336 & ($-$1.27473, $-$1.27200) \\

& 2.99902 & $-$1.27336 & ($-$1.27467, $-$1.27207) \\

\\

1.13169 & 3.04766 & $-$1.27422 & ($-$1.27863, $-$1.26983) \\

& 3.04837 & $-$1.27425 & ($-$1.27562, $-$1.27290) \\

& 3.04838 & $-$1.27425 & ($-$1.27556, $-$1.27296) \\

\\

1.13301 & 3.10580 & $-$1.27511 & ($-$1.27952, $-$1.27071) \\

& 3.10677 & $-$1.27514 & ($-$1.27652, $-$1.27379) \\

& 3.10678 & $-$1.27514 & ($-$1.27645, $-$1.27386) \\
 
\end{tabular}

\end{ruledtabular}

\end{table}}

\clearpage

\clearpage

\clearpage

\clearpage

{\squeezetable

\begin{table}

\caption{Values of $\chi^2$ at sample points in the
$(\lambda,R)$-plane, corresponding to Table~\ref{tabla11}. The
upper, middle, and lower entries have the same meaning as in this
table.
\label{tabla2}}

\begin{ruledtabular}

\begin{tabular}{lddddddddd}

$R'(10^{-3}{\rm s}^{-1})\backslash\lambda'$ & -1.2744 & -1.2720 &
-1.2696 & -1.2672 & -1.2648 & -1.2624 & -1.2600 & -1.2576 &
-1.2552 \\

\hline

1.12509 & 14.82 & 9.43 & 5.67 & 3.52 & 3.00 & 4.10 & 6.82
& 11.17 & 17.14 \\

 & 127.66 & 72.82 & 32.96 & 9.31 & 3.25 & 16.31 & 50.21 &
106.90 & 188.54 \\

 & 139.40 & 79.79 & 36.14 & 10.03 & 3.28 & 18.02 & 56.75 &
122.43 & 218.63 \\

 \\
 
1.12641 & 12.60 & 7.81 & 4.65 & 3.11 & 3.18 & 4.89 & 8.21
& 13.16 & 19.73 \\

 & 105.461 & 56.10 & 22.15 & 4.89 & 5.75 & 26.35 & 68.46 &
134.12 & 225.60 \\

 & 115.31 & 61.51 & 24.22 & 5.11 & 6.10 & 29.43 & 77.78
& 154.30 & 262.84 \\

 \\
 
1.12773 & 10.61 & 6.43 & 3.86 & 2.92 & 3.60 & 5.90 & 9.83
& 15.38 & 22.56 \\

 & 85.29 & 41.55 & 13.67 & 2.98 & 10.99 & 39.35 & 89.94 &
164.86 & 266.52 \\

 & 93.36 & 45.56 & 14.86 & 2.99 & 12.00 & 44.28 & 102.65 &
190.52 & 312.02 \\

 \\
 
1.12905 & 8.86 & 5.27 & 3.31 & 2.97 & 4.24 & 7.15 & 11.68
& 17.83 & 25.61 \\

 & 67.19 & 29.23 & 7.59 & 3.67 & 19.03 & 55.40 & 114.73 &
199.24 & 311.41 \\

 & 73.63 & 32.02 & 8.12 & 3.77 & 21.11 & 62.70 & 131.55 &
231.30 & 366.41 \\

 \\
 
1.13037 & 7.34 & 4.35 & 2.99 & 3.24 & 5.12 & 8.63 & 13.76
& 20.51 & 28.90 \\

 & 51.22 & 19.19 & 3.98 & 7.03 & 29.96 & 74.59 & 142.95 &
237.36 & 360.42 \\

 & 56.16 & 20.96 & 4.11 & 7.54 & 33.55 & 84.84 & 164.63 &
276.85 & 426.31 \\

 \\
 
1.13169 & 6.05 & 3.66 & 2.89 & 3.75 & 6.23 & 10.34 & 16.07
& 23.43 & 32.41 \\

 & 37.43 & 11.51 & 2.90 & 13.13 & 43.87 & 97.02 & 174.71 &
279.35 & 413.67 \\

 & 41.04 & 12.46 & 2.90 & 14.42 & 49.45 & 110.86 & 202.10
& 327.41 & 492.01 \\

 \\
 
1.13301 & 4.99 & 3.20 & 3.03 & 4.49 & 7.57 & 12.28 & 18.61
& 26.57 & 36.16 \\

 & 25.88 & 6.22 & 4.42 & 22.04 & 60.83 & 122.78 & 210.11 &
325.34 & 471.33 \\

 & 28.33 & 6.60 & 4.61 & 24.53 & 68.94 & 140.91 & 244.16 &
383.24 & 563.83 \\

 \\

\end{tabular}

\end{ruledtabular}

\end{table}}

\clearpage

{\squeezetable

\begin{table}

\caption{Values of $\chi^2$ at sample points in the
$(\lambda,R)$-plane, corresponding to Table~\ref{tabla14}. The
upper, middle, and lower entries have the same meaning as in this
table. \label{tabla24}}

\begin{ruledtabular}

\begin{tabular}{lddddddddd}

$R'(10^{-3}{\rm s}^{-1})\backslash\lambda'$ & -1.2744 & -1.2720 &
-1.2696 & -1.2672 & -1.2648 & -1.2624 & -1.2600 & -1.2576 &
-1.2552 \\

\hline

1.12509 & 9.64 & 5.78 & 3.54 & 2.91 & 3.90 & 6.50 & 10.71 & 16.55
& 24.00 \\

 & 71.89 & 33.20 & 9.82 & 3.01 & 14.14 & 44.77 & 96.62 & 171.64 &
272.01 \\

 & 78.36 & 36.21 & 10.55 & 3.02 & 15.49 & 50.14 & 109.54 & 196.75
& 315.44 \\

 \\
 
1.12641 & 8.01 & 4.75 & 3.11 & 3.08 & 4.66 & 7.86 & 12.68 & 19.11
& 27.16 \\

 & 55.70 & 22.61 & 5.26 & 4.97 & 23.16 & 61.46 & 121.66 & 205.79
& 316.13 \\

 & 60.76 & 24.60 & 5.52 & 5.20 & 25.65 & 69.16 & 138.48 & 236.86
& 368.27 \\

 \\
 
1.12773 & 6.61 & 3.95 & 2.91 & 3.48 & 5.66 & 9.48 & 14.87 & 21.91
& 30.56 \\

 & 41.58 & 14.23 & 3.08 & 9.49 & 34.96 & 81.16 & 149.97 & 243.51
& 364.16 \\

 & 45.36 & 15.40 & 3.10 & 10.24 & 38.99 & 91.72 & 171.38 & 281.45
& 426.20 \\

 \\
 
1.12905 & 5.45 & 3.38 & 2.94 & 4.10 & 6.88 & 11.28 & 17.30 &
24.93 & 34.19 \\

 & 29.58 & 8.12 & 3.35 & 16.65 & 49.61 & 103.95 & 181.65 &
284.90 & 416.21 \\

 & 32.24 & 8.67 & 3.40 & 18.31 & 55.63 & 117.96 & 208.41 &
330.73 & 489.51 \\

 \\
 
1.13037 & 4.51 & 3.05 & 3.20 & 4.96 & 8.34 & 13.34 & 19.95 &
28.19 & 38.04 \\

 & 19.74 & 4.34 & 6.12 & 26.52 & 67.19 & 129.93 & 216.79 &
330.09 & 472.41 \\

 & 21.45 & 4.50 & 6.49 & 29.44 & 75.69 & 148.02 & 249.75 &
384.91 & 558.47 \\

 \\
 
1.13169 & 3.81 & 2.94 & 3.69 & 6.05 & 10.03 & 15.62 & 22.84 &
31.67 & 42.13 \\

 & 12.12 & 2.95 & 11.46 & 39.18 & 87.78 & 159.19 & 255.52 &
379.19 & 532.91 \\

 & 13.07 & 2.95 & 12.47 & 43.76 & 99.30 & 182.06 & 295.60 &
444.25 & 633.39 \\

 \\
 
1.13301 & 3.33 & 3.06 & 4.41 & 7.37 & 11.94 & 18.14 & 25.95 &
35.39 & 46.44 \\

 & 6.77 & 4.00 & 19.45 & 54.69 & 111.49 & 191.82 & 297.93 &
432.33 & 597.85 \\

 & 7.14 & 4.12 & 21.44 & 61.36 & 126.61 & 220.27 & 346.17 &
509.01 & 714.63 \\

 \\

\end{tabular}

\end{ruledtabular}

\end{table}}

\clearpage

{\squeezetable

\begin{table}

\caption{Values of $\chi^2$ at sample points in the
$(\lambda,R)$-plane, corresponding to Table~\ref{tabla17}. The
upper, middle, and lower entries have the same meaning as in this
table.\label{tabla27}}

\begin{ruledtabular}

\begin{tabular}{lddddddddd}

$R'(10^{-3}{\rm s}^{-1})\backslash\lambda'$ & -1.2744 & -1.2720 &
-1.2696 & -1.2672 & -1.2648 & -1.2624 & -1.2600 & -1.2576 &
-1.2552 \\

\hline

1.12509 & 5.87 & 3.58 & 2.90 & 3.82 & 6.34 & 10.48 & 16.22 &
23.57 & 32.53 \\

 & 32.54 & 9.91 & 2.94 & 12.87 & 41.11 & 89.20 & 158.89 & 252.12
& 371.07 \\

 & 35.31 & 10.61 & 2.95 & 13.99 & 45.67 & 100.23 & 180.31 &
289.04 & 430.22 \\

 \\
 
1.12641 & 4.85 & 3.15 & 3.06 & 4.58 & 7.70 & 12.43 & 18.77 &
26.72 & 36.28 \\

 & 22.34 & 5.43 & 4.61 & 21.19 & 56.61 & 112.51 & 190.68 &
293.15 & 422.22 \\

 & 24.20 & 5.68 & 4.79 & 23.28 & 63.18 & 126.87 & 217.16 &
337.40 & 491.66 \\

 \\
 
1.12773 & 4.05 & 2.95 & 3.46 & 5.57 & 9.29 & 14.62 & 21.56 &
30.11 & 40.27 \\

 & 14.25 & 3.20 & 8.70 & 32.11 & 74.93 & 138.86 & 225.78 &
337.80 & 477.30 \\

 & 15.35 & 3.22 & 9.33 & 35.52 & 83.95 & 157.14 & 258.08 &
390.36 & 558.36 \\

 \\
 
1.12905 & 3.49 & 2.99 & 4.09 & 6.80 & 11.11 & 17.04 & 24.57 &
33.72 & 44.48 \\

 & 8.30 & 3.27 & 15.27 & 45.70 & 96.15 & 168.35 & 264.29 &
386.16 & 536.45 \\

 & 8.83 & 3.31 & 16.65 & 50.83 & 108.12 & 191.19 & 303.24 &
448.13 & 630.57 \\

 \\
 
1.13037 & 3.16 & 3.25 & 4.95 & 8.25 & 13.16 & 19.68 & 27.82 &
37.56 & 48.92 \\

 & 4.56 & 5.72 & 24.39 & 62.05 & 120.34 & 201.07 & 306.31 &
438.367 & 599.80 \\

 & 4.72 & 6.01 & 26.85 & 69.315 & 135.80 & 229.17 & 352.84 &
510.95 & 708.59 \\

 \\
 
1.13169 & 3.05 & 3.74 & 6.03 & 9.93 & 15.44 & 22.56 & 31.29 &
41.63 & 53.56 \\

 & 3.08 & 10.58 & 36.12 & 81.22 & 147.58 & 237.12 & 351.96 &
494.51 & 667.49 \\

 & 3.08 & 11.42 & 40.02 & 91.07 & 167.13 & 271.25 & 407.07 &
579.06 & 792.75 \\

 \\
 
1.13301 & 3.18 & 4.46 & 7.35 & 11.84 & 17.95 & 25.66 & 34.99 &
45.93 & 58.48 \\

 & 3.91 & 17.94 & 50.54 & 103.30 & 177.98 & 276.58 & 401.33 &
554.75 & 739.65 \\

 & 4.00 & 19.62 & 56.26 & 116.23 & 202.26 & 317.60 & 466.15 &
652.74 & 883.38 \\

 \\

\end{tabular}

\end{ruledtabular}

\end{table}}

\clearpage

{\squeezetable

\begin{table}

\caption{In the top part we give the individual contributions to
$\chi^2$ of Eq.~(\ref{chi2}) and total $\chi^2$ at the border
points of the $90\%$ CL ranges of $\lambda$ of the middle row of
Table~\ref{tabla14}. The upper, middle, and lower entries in each
row have the same meaning as in this table. In the lower part we
assume that $\sigma_{R}$ and $\sigma_ {\lambda}$ are cut to
$1/2$, $1/5$, and $1/7$ and correspond to the second, third, and
fourth entries in each row.
\label{tablavii}}

\begin{ruledtabular}

\begin{tabular}{cccccccc}

$\lambda'$ & $\chi^2(R')$ & $\chi^2(\lambda')$ & $\chi^2(V_{\rm us})$ &
$\chi^2(B_0)$ & $\chi^2(a_0)$ & $\chi^2(V_{\rm ub})$ &  $\chi^2$ \\

\hline

$-$1.27452 & 0.29217 & 2.17752 & 0.21557 & 2.59159 & 0.32058 & $10^{-6}$ &
5.59743 \\

$-$1.27147 & 0.03195 & 0.23217 & 2.40786 & 2.56871 & 0.35276 & 0.00002 &
5.59347 \\

$-$1.27140 & 0.00035 & 0.00257 & 2.66590 & 2.56582 & 0.35694 & 0.00002 &
5.59160 \\

\\

$-$1.26575 & 0.30545 & 2.17991 & 0.22051 & 2.69656 & 0.19360 & $10^{-6}$ &
5.59603 \\

$-$1.26882 & 0.03432 & 0.24672 & 2.42558 & 2.71815 & 0.17160 & 0.00002 &
5.59639 \\

$-$1.26889 & 0.00038 & 0.00275 & 2.70112 & 2.72118 & 0.16862 & 0.00002 &
5.59408 \\

\\

\hline

\\

$-$1.27452 & 0.29217 & 2.17752 & 0.21557 & 2.59159 & 0.32058 & $10^{-6}$ &
5.59743 \\

$-$1.27259 & 0.23756 & 1.74066 & 0.70608 & 2.59012 & 0.32259 & $10^{-5}$ &
5.59702 \\

$-$1.27166 & 0.09891 & 0.71964 & 1.85638 & 2.57525 & 0.34339 & $10^{-5}$ &
5.59360 \\

$-$1.27154 & 0.05925 & 0.43072 & 2.18460 & 2.57129 & 0.34904 & 0.00002 &
5.59492 \\

$-$1.27147 & 0.03195 & 0.23217 & 2.40786 & 2.56871 & 0.35276 & 0.00002 &
5.59347 \\

\\

$-$1.26575 & 0.30545 & 2.17991 & 0.22051 & 2.69656 & 0.19360 & $10^{-6}$ &
5.59603 \\

$-$1.26769 & 0.24571 & 1.75902 & 0.70465 & 2.69705 & 0.19307 & $10^{-5}$ &
5.59951 \\

$-$1.26864 & 0.10440 & 0.74983 & 1.85193 & 2.71146 & 0.17827 & $10^{-5}$ &
5.59591 \\

$-$1.26876 & 0.06321 & 0.45421 & 2.19225 & 2.71550 & 0.17422 & 0.00002 &
5.59942 \\

$-$1.26882 & 0.03432 & 0.24672 & 2.42558 & 2.71815 & 0.17160 & 0.00002 &
5.59639 \\
 
\end{tabular}

\end{ruledtabular}

\end{table}}

\clearpage

{\squeezetable

\begin{table}

\caption{This table corresponds to Table~\ref{tabla11}, except
that now it is assumed that $\sigma_{V_{\rm us}}$ is cut to
$1/10$, namely, $\sigma_{V_{\rm us}}=0.00021$.\label{tabla11v9}}

\begin{ruledtabular}

\begin{tabular}{cccc}

$R'$ & $\chi^2_0$ & $\lambda_0$ & $\lambda'$ \\

\hline

1.12509 & 2.97483 & $-$1.26521 & ($-$1.26942, $-$1.26100) \\

& 2.97522 & $-$1.26520 & ($-$1.26563, $-$1.26476)  \\

& 2.97523 & $-$1.26520 & ($-$1.26532, $-$1.26507) \\

\\

1.12641 & 2.93950 & $-$1.26611 & ($-$1.27032, $-$1.26190) \\

& 2.93974 & $-$1.26609 & ($-$1.26653, $-$1.26565) \\

& 2.93974 & $-$1.26609 & ($-$1.26622, $-$1.26596) \\

\\

1.12773 & 2.91364 & $-$1.26700 & ($-$1.27121, $-$1.26279) \\

& 2.91375 & $-$1.26699 & ($-$1.26743, $-$1.26655) \\

& 2.91375 & $-$1.26699 & ($-$1.26712, $-$1.26686) \\

\\

1.12905 & 2.89718 & $-$1.26789 & ($-$1.27210, $-$1.26368) \\

& 2.89721 & $-$1.26788 & ($-$1.26832, $-$1.26745) \\

& 2.89721 & $-$1.26788 & ($-$1.26801, $-$1.26775) \\

\\

1.13037 & 2.89005 & $-$1.26878 & ($-$1.27299, $-$1.26457) \\

& 2.89005 & $-$1.26878 & ($-$1.26922, $-$1.26834) \\

& 2.89005 & $-$1.26878 & ($-$1.26891, $-$1.26865) \\

\\

1.13169 & 2.89221 & $-$1.26967 & ($-$1.27388, $-$1.26546) \\

& 2.89222 & $-$1.26967 & ($-$1.27011, $-$1.26923) \\

& 2.89222 & $-$1.26967 & ($-$1.26980, $-$1.26954) \\

\\

1.13301 & 2.90358 & $-$1.27056 & ($-$1.27477, $-$1.26635) \\

& 2.90364 & $-$1.27056 & ($-$1.27100, $-$1.27013) \\

& 2.90364 & $-$1.27056 & ($-$1.27069, $-$1.27044) \\
 
\end{tabular}

\end{ruledtabular}

\end{table}}

\clearpage

{\squeezetable

\begin{table}

\caption{This table corresponds to Table~\ref{tabla14}, except
that now it is assumed that $\sigma_{V_{\rm us}}$ is cut to
$1/10$, namely, $\sigma_{V_{\rm us}}=0.00021$.\label{tabla14v9}}

\begin{ruledtabular}

\begin{tabular}{cccc}

$R'$ & $\chi^2_0$ & $\lambda_0$ & $\lambda'$ \\

\hline

1.12509 & 2.90394 & $-$1.26746 & ($-$1.27167, $-$1.26325) \\

& 2.90400 & $-$1.26745 & ($-$1.26789, $-$1.26701)  \\

& 2.90402 & $-$1.26745 & ($-$1.26759, $-$1.26732) \\

\\

1.12641 & 2.89230 & $-$1.26835 & ($-$1.27256, $-$1.26414) \\

& 2.89231 & $-$1.26835 & ($-$1.26879, $-$1.26791) \\

& 2.89231 & $-$1.26835 & ($-$1.26848, $-$1.26822) \\

\\

1.12773 & 2.89003 & $-$1.26925 & ($-$1.27346, $-$1.26504) \\

& 2.89003 & $-$1.26925 & ($-$1.26969, $-$1.26881) \\

& 2.89009 & $-$1.26925 & ($-$1.26938, $-$1.26911) \\

\\

1.12905 & 2.89705 & $-$1.27014 & ($-$1.27435, $-$1.26593) \\

& 2.89709 & $-$1.27014 & ($-$1.27058, $-$1.26970) \\

& 2.89710 & $-$1.27014 & ($-$1.27027, $-$1.27001) \\

\\

1.13037 & 2.91333 & $-$1.27103 & ($-$1.27524, $-$1.26682) \\

& 2.91344 & $-$1.27104 & ($-$1.27148, $-$1.27060) \\

& 2.91350 & $-$1.27104 & ($-$1.27117, $-$1.27091) \\

\\

1.13169 & 2.93878 & $-$1.27192 & ($-$1.27613, $-$1.26771) \\

& 2.93901 & $-$1.27193 & ($-$1.27237, $-$1.27149) \\

& 2.93903 & $-$1.27193 & ($-$1.27207, $-$1.27180) \\

\\

1.13301 & 2.97336 & $-$1.27281 & ($-$1.27702, $-$1.26860) \\

& 2.97375 & $-$1.27283 & ($-$1.27327, $-$1.27239) \\

& 2.97380 & $-$1.27283 & ($-$1.27296, $-$1.27269) \\
 
\end{tabular}

\end{ruledtabular}

\end{table}}

\clearpage

{\squeezetable

\begin{table}

\caption{This table corresponds to Table~\ref{tabla17}, except
that now it is assumed that $\sigma_{V_{\rm us}}$ is cut to
$1/10$, namely, $\sigma_{V_{\rm us}}=0.00021$.\label{tabla17v9}}

\begin{ruledtabular}

\begin{tabular}{cccc}

$R'$ & $\chi^2_0$ & $\lambda_0$ & $\lambda'$ \\

\hline

1.12509 & 2.89311 & $-$1.26978 & ($-$1.27399, $-$1.26557) \\

& 2.89313 & $-$1.26978 & ($-$1.27022, $-$1.26934)  \\

& 2.89314 & $-$1.26978 & ($-$1.26992, $-$1.26965) \\

\\

1.12641 & 2.90572 & $-$1.27067 & ($-$1.27488, $-$1.26646) \\

& 2.90572 & $-$1.27067 & ($-$1.27112, $-$1.27024) \\

& 2.90587 & $-$1.27068 & ($-$1.27082, $-$1.27054) \\

\\

1.12773 & 2.92760 & $-$1.27157 & ($-$1.27578, $-$1.26736) \\

& 2.92778 & $-$1.27158 & ($-$1.27202, $-$1.27114) \\

& 2.92791 & $-$1.27158 & ($-$1.27171, $-$1.27144) \\

\\

1.12905 & 2.95868 & $-$1.27246 & ($-$1.27667, $-$1.26825) \\

& 2.95900 & $-$1.27248 & ($-$1.27292, $-$1.27203) \\

& 2.95905 & $-$1.27248 & ($-$1.27261, $-$1.27234) \\

\\

1.13037 & 2.99890 & $-$1.27335 & ($-$1.27756, $-$1.26914) \\

& 2.99941 & $-$1.27337 & ($-$1.27381, $-$1.27293) \\

& 2.99942 & $-$1.27337 & ($-$1.27351, $-$1.27324) \\

\\

1.13169 & 3.04822 & $-$1.27424 & ($-$1.27845, $-$1.27003) \\

& 3.04894 & $-$1.27427 & ($-$1.27471, $-$1.27383) \\

& 3.04897 & $-$1.27427 & ($-$1.27441, $-$1.27413) \\

\\

1.13301 & 3.10656 & $-$1.27514 & ($-$1.27935, $-$1.27093) \\

& 3.10754 & $-$1.27516 & ($-$1.27560, $-$1.27472) \\

& 3.10760 & $-$1.27516 & ($-$1.27530, $-$1.27503) \\
 
\end{tabular}

\end{ruledtabular}

\end{table}}

\clearpage

{\squeezetable

\begin{table}

\caption{This table corresponds to Table~\ref{tabla2}, except
that now it is assumed that $\sigma_{V_{\rm us}}$ is cut to
$1/10$, namely, $\sigma_{V_{\rm us}}=0.00021$.
\label{tabla2v9}}

\begin{ruledtabular}

\begin{tabular}{lddddddddd}

$R'(10^{-3}{\rm s}^{-1})\backslash\lambda'$ & -1.2744 & -1.2720 &
-1.2696 & -1.2672 & -1.2648 & -1.2624 & -1.2600 & -1.2576 &
-1.2552 \\

\hline

1.12509 & 15.83 & 9.99 & 5.90 & 3.58 & 3.00 & 4.18 & 7.12
& 11.80 & 18.25 \\

 & 1193.11 & 653.49 & 275.56 & 59.42 & 5.18 & 112.94 & 382.79 &
814.84 & 1409.18 \\

 & 12460.00 & 6988.55 & 3008.19 & 642.41 & 28.65 & 1320.42 &
4689.62 & 10329.40 & 18457.20 \\

 \\
 
1.12641 & 13.42 & 8.23 & 4.80 & 3.12 & 3.20 & 5.03 & 8.62
& 13.96 & 21.06 \\

 & 972.71 & 493.42 & 175.88 & 20.18 & 26.44 & 194.73 & 525.18 &
1017.87 & 1672.91 \\

 & 10241.20 & 5316.84 & 1927.06 & 200.14 & 279.14 & 2323.96 &
6513.91 & 13050.60 & 22161.40 \\

 \\
 
1.12773 & 11.26 & 6.73 & 3.95 & 2.92 & 3.65 & 6.13 & 10.37
& 16.37 & 24.12 \\

 & 774.96 & 355.99 & 98.82 & 3.54 & 70.27 & 299.09 & 690.12 &
1243.44 & 1959.15 \\

 & 8224.81 & 3862.50 & 1079.81 & 10.20 & 802.576 & 3623.61 &
8660.38 & 16123.40 & 26250.40 \\

 \\
 
1.12905 & 9.36 & 5.47 & 3.34 & 2.97 & 4.35 & 7.49 & 12.38
& 19.03 & 27.43 \\

 & 599.83 & 241.16 & 44.34 & 9.46 & 136.64 & 425.97 & 877.56 &
1491.49 & 2267.86 \\

 & 6415.87 & 2630.99 & 472.56 & 79.42 & 1606.66 & 5228.04 &
11138.80 & 19559.00 & 30737.00 \\

 \\
 
1.13037 & 7.71 & 4.47 & 2.99 & 3.27 & 5.30 & 9.09 & 14.63
& 21.93 & 30.99 \\

 & 447.28 & 148.89 & 12.40 & 37.91 & 225.53 & 575.35 & 1087.47 &
1761.99 & 2599.01 \\

 & 4819.44 & 1627.99 & 111.661 & 414.94 & 2699.41 & 7146.31 &
13959.50 & 23368.80 & 35634.30 \\

 \\
 
1.13169 & 6.30 & 3.72 & 2.89 & 3.82 & 6.51 & 10.95 & 17.14
& 25.09 & 34.80 \\

 & 317.26 & 79.14 & 2.96 & 88.84 & 336.88 & 747.17 & 1319.81 &
2054.91 & 2952.55 \\

 & 3440.83 & 859.42 & 3.73 & 1024.18 & 4089.20 & 9387.86 &
17133.10 & 27565.20 & 40956.00 \\

 \\
 
1.13301 & 5.15 & 3.22 & 3.04 & 4.62 & 7.96 & 13.05 & 19.90
& 28.50 & 38.86 \\

 & 209.76 & 31.88 & 16.00 & 162.23 & 470.67 & 941.41 & 1574.55 &
2370.20 & 3328.45 \\

 & 2285.50 & 331.40 & 155.64 & 1914.89 & 5784.74 & 11962.50 &
20670.70 & 32160.60 & 46716.70 \\

 \\

\end{tabular}

\end{ruledtabular}

\end{table}}

{\squeezetable

\begin{table}

\caption{This table corresponds to Table~\ref{tabla24}, except
that now it is assumed that $\sigma_{V_{\rm us}}$ is cut to
$1/10$, namely, $\sigma_{V_{\rm us}}=0.00021$.\label{tabla24v9}}

\begin{ruledtabular}

\begin{tabular}{lddddddddd}

$R'(10^{-3}{\rm s}^{-1})\backslash\lambda'$ & -1.2744 & -1.2720 &
-1.2696 & -1.2672 & -1.2648 & -1.2624 & -1.2600 & -1.2576 &
-1.2552 \\

\hline

1.12509 & 10.24 & 6.04 & 3.60 & 2.91 & 3.98 & 6.80 & 11.38 &
17.70 & 25.78 \\

 & 677.51 & 292.00 & 67.39 & 3.79 & 101.31 & 360.05 & 780.13 &
1361.63 & 2104.68 \\

 & 6885.08 & 3027.71 & 695.41 & 12.72 & 1118.13 & 4166.06 &
9329.22 & 16801.30 & 26800.00 \\

 \\
 
1.12641 & 8.46 & 4.92 & 3.13 & 3.09 & 4.81 & 8.29 & 13.52 & 20.50
& 29.23 \\

 & 514.57 & 189.18 & 24.75 & 21.38 & 179.19 & 498.27 & 978.73 &
1620.68 & 2424.23 \\

 & 5267.30 & 1969.07 & 239.73 & 208.68 & 2019.96 & 5834.38 &
11832.00 & 20214.80 & 31210.20 \\

 \\
 
1.12773 & 6.94 & 4.05 & 2.91 & 3.53 & 5.90 & 10.03 & 15.91 &
23.54 & 32.93 \\

 & 374.19 & 108.91 & 4.64 & 61.48 & 279.55 & 658.96 & 1199.79 &
1902.17 & 2766.19 \\

 & 3855.82 & 1131.72 & 21.99 & 661.14 & 3199.04 & 7803.19 &
14661.30 & 23984.30 & 36009.60 \\

 \\
 
1.12905 & 5.66 & 3.43 & 2.94 & 4.21 & 7.24 & 12.02 & 18.55 &
26.84 & 36.88 \\

 & 256.34 & 51.14 & 7.02 & 124.06 & 402.38 & 842.09 & 1443.28 &
2206.07 & 3130.55 \\

 & 2655.59 & 521.18 & 48.36 & 1376.99 & 4663.08 & 10081.20 &
17827.00 & 28120.90 & 41210.90 \\

 \\
 
1.13037 & 4.64 & 3.06 & 3.22 & 5.15 & 8.82 & 14.26 & 21.44 &
30.38 & 41.08 \\

 & 160.97 & 15.85 & 31.85 & 209.07 & 547.63 & 1047.62 & 1709.15 &
2532.33 & 3517.26 \\

 & 1671.75 & 143.163 & 325.22 & 2363.39 & 6420.14 & 12677.40 &
21339.52 & 32636.20 & 46827.20 \\

 \\
 
1.13169 & 3.88 & 2.94 & 3.76 & 6.33 & 10.66 & 16.75 & 24.59 &
34.18 & 45.53 \\

 & 88.07 & 3.00 & 79.11 & 316.49 & 715.26 & 1275.52 & 1997.37 &
2880.92 & 3926.27 \\

 & 909.64 & 3.62 & 859.21 & 3627.8 & 8478.60 & 15601.30 &
25209.20 & 37542.40 & 52872.10 \\

 \\
 
1.13301 & 3.36 & 3.07 & 4.54 & 7.77 & 12.75 & 19.49 & 27.98 &
38.23 & 50.23 \\

 & 37.58 & 12.55 & 148.75 & 446.27 & 905.24 & 1525.75 & 2307.91 &
3251.81 & 4357.57 \\

 & 374.793 & 108.74 & 1657.27 & 5177.98 & 10847.20 & 18862.80 &
29447.40 & 42852.00 & 59360.20 \\

 \\

\end{tabular}

\end{ruledtabular}

\end{table}}

{\squeezetable

\begin{table}

\caption{This table corresponds to Table~\ref{tabla27}, except
that now it is assumed that $\sigma_{V_{\rm us}}$ is cut to
$1/10$, namely, $\sigma_{V_{\rm us}}=0.00021$.
\label{tabla27v9}}

\begin{ruledtabular}

\begin{tabular}{lddddddddd}

$R'(10^{-3}{\rm s}^{-1})\backslash\lambda'$ & -1.2744 & -1.2720 &
-1.2696 & -1.2672 & -1.2648 & -1.2624 & -1.2600 & -1.2576 &
-1.2552 \\

\hline

1.12509 & 6.14 & 3.64 & 2.90 & 3.91 & 6.67 & 11.18 & 17.44 &
25.46 & 35.24 \\

 & 299.31 & 71.28 & 3.35 & 95.64 & 348.25 & 761.30 & 1334.90 &
2069.16 & 2964.19 \\

 & 2951.38 & 700.34 & 7.72 & 999.09 & 3813.89 & 8607.42 &
15553.10 & 24845.10 & 36701.30 \\

 \\
 
1.12641 & 5.02 & 3.17 & 3.08 & 4.74 & 8.16 & 13.32 & 20.24 &
28.92 & 39.34 \\

 & 195.23 & 27.11 & 19.15 & 171.45 & 484.14 & 957.31 & 1591.10 &
2385.60 & 3340.92 \\

 & 1932.00 & 251.88 & 174.37 & 1829.89 & 5363.41 & 10936.60 &
18730.00 & 28946.10 & 41812.50 \\

 \\
 
1.12773 & 4.15 & 2.96 & 3.52 & 5.83 & 9.90 & 15.72 & 23.29 &
32.62 & 43.70 \\

 & 113.63 & 5.40 & 57.38 & 269.69 & 642.43 & 1175.72 & 1869.67 &
2724.40 & 3740.00 \\

 & 1122.64 & 28.56 & 582.94 & 2921.28 & 7194.36 & 13570.40 &
22237.80 & 33407.50 & 47317.20 \\

 \\
 
1.12905 & 3.53 & 2.99 & 4.20 & 7.17 & 11.89 & 18.37 & 26.60 &
36.58 & 48.31 \\

 & 54.46 & 6.11 & 118.02 & 390.31 & 823.09 & 1416.48 & 2170.58 &
3085.51 & 4161.37 \\

 & 528.29 & 35.93 & 1239.62 & 4280.18 & 9314.50 & 16517.80 &
26086.30 & 38240.40 & 53227.90 \\

 \\
 
1.13037 & 3.17 & 3.28 & 5.14 & 8.76 & 14.14 & 21.27 & 30.15 &
40.78 & 53.17 \\

 & 17.69 & 29.20 & 201.02 & 533.28 & 1026.09 & 1679.56 & 2493.79
& 3468.91 & 4605.00 \\

 & 154.13 & 279.75 & 2150.84 & 5913.79 & 11731.90 & 19787.70 &
30285.80 & 43456.20 & 59557.70 \\

 \\
 
1.13169 & 3.05 & 3.82 & 6.33 & 10.61 & 16.63 & 24.41 & 33.95 &
45.24 & 58.28 \\

 & 3.29 & 74.64 & 306.36 & 698.57 & 1251.39 & 1964.91 & 2839.26 &
3874.55 & 5070.87 \\

 & 5.55 & 766.02 & 3323.28 & 7829.59 & 14454.90 & 23389.60 &
34846.90 & 49067.00 & 66320.30 \\

 \\
 
1.13301 & 3.19 & 4.60 & 7.77 & 12.70 & 19.38 & 27.81 & 38.00 &
49.95 & 63.64 \\

 & 11.22 & 142.39 & 433.99 & 886.14 & 1498.94 & 2272.51 & 3206.96
& 4302.40 & 5558.92 \\

 & 88.11 & 1500.95 & 4763.89 & 10035.40 & 17492.40 & 27333.30 &
39780.70 & 55085.40 & 73529.80 \\

 \\

\end{tabular}

\end{ruledtabular}

\end{table}}

\clearpage

\clearpage

\clearpage

\clearpage

\clearpage

{\squeezetable

\begin{table}

\caption{Values of $V_{\rm ud}$ and $V_{\rm us}$ assuming the
central values of $R$ and $\lambda$ are within the small regions
displayed in Fig.~\ref{fig_num7}. The $-$ and $+$ indices
correspond to the left- and right-hand horizontal
border points of each of the larger $90\%$ CL regions
in this figure. \label{tablaxiv}}

\begin{ruledtabular}

\begin{tabular}{ccc}

$\lambda$ & $V_{\rm ud}$ & $V_{\rm us}$ \\

\hline
 
$\lambda^-_{\rm A}=-1.2770$ & $0.96984\pm 0.00020$ &
$0.24372\pm 0.00078$ \\
 
$\lambda_{\rm A}=-1.2739$ & $0.97180\pm 0.00020$ &
$0.23575\pm 0.00081$ \\
 
$\lambda^+_{\rm A}=-1.2708$ & $0.97378\pm 0.00020$ &
$0.22745\pm 0.00085$ \\

\\
 
$\lambda^-_{\rm ALYB}=-1.2720$ & $0.97303\pm 0.00016$ &
$0.23062\pm 0.00066$ \\
 
$\lambda_{\rm ALYB}=-1.2695$ & $0.97460\pm 0.00016$ &
$0.22393\pm 0.00068$ \\
 
$\lambda^+_{\rm ALYB}=-1.2670$ & $0.97616\pm 0.00016$ &
$0.21699\pm 0.00070$ \\

\\
 
$\lambda^-_{\rm LYB}=-1.2663$ & $0.97661\pm 0.00025$ &
$0.2150\pm 0.0011$ \\
 
$\lambda_{\rm LYB}=-1.2624$ & $0.97913\pm 0.00025$ &
$0.2032\pm 0.0012$ \\
 
$\lambda^+_{\rm LYB}=-1.2585$ & $0.98166\pm 0.00025$ &
$0.1906\pm 0.0013$ \\
 
\end{tabular}

\end{ruledtabular}

\end{table}}


\begin{thebibliography}{}
\bibitem{review}
A review and original references are found
in Ref.~\cite{pdg06}.
\end{thebibliography}
\end{document}